%% file: helicity.tex
\documentclass[traditabstract]{aa}

\usepackage{amsmath}
\usepackage{graphicx}
\usepackage{multirow}
\usepackage{txfonts}
\usepackage[authoryear]{natbib}
\usepackage{hyperref,ifthen}

\bibpunct{(}{)}{;}{a}{}{,}

\begin{document}

\title{Probing magnetic helicity with synchrotron radiation and Faraday rotation}
\author{N. Oppermann \and H. Junklewitz \and G. Robbers \and T.A. En\ss{}lin}
\institute{Max Planck Institute for Astrophysics, Karl-Schwarzschild-Str. 1, 85741 Garching, Germany\\ \email{niels@mpa-garching.mpg.de}}
\date{Received 06 Aug. 2010 / Accepted 01 Feb. 2011}
\abstract{
We present a first application of the recently proposed \textit{LITMUS} test for magnetic helicity, as well as a thorough study of its applicability under different circumstances. In order to apply this test to the galactic magnetic field, the newly developed \textit{critical filter} formalism is used to produce an all-sky map of the Faraday depth. The test does not detect helicity in the galactic magnetic field. To understand the significance of this finding, we made an applicability study, showing that a definite conclusion about the absence of magnetic helicity in the galactic field has not yet been reached. This study is conducted by applying the test to simulated observational data. We consider simulations in a flat sky approximation and all-sky simulations, both with assumptions of constant electron densities and realistic distributions of thermal and cosmic ray electrons. Our results suggest that the \textit{LITMUS} test does indeed perform very well in cases where constant electron densities can be assumed, both in the flat-sky limit and in the galactic setting. Non-trivial distributions of thermal and cosmic ray electrons, however, may complicate the scenario to the point where helicity in the magnetic field can escape detection.
}

\titlerunning{Probing Magnetic Helicity}
\authorrunning{N. Oppermann et al.}
\maketitle

\section{Introduction}

Helicity is of utmost interest in the study of astrophysical magnetism. Mean field theories for turbulent dynamos operating in the galactic interstellar medium have been successful in explaining how the observed magnetic field strengths are maintained \citep[e.g.][]{subramanian-2002}. These theories predict that helicity is present on small scales in interstellar magnetic fields. Observationally detecting or excluding helicity in these fields would therefore either strongly suggest that these theories are valid or indicate that there are some flaws in them.

However, since helicity is a quantity that describes the three-dimensional structure of a magnetic field and most observation techniques produce at best two-dimensional images leading to an informational deficit, it has thus far largely eluded observers. Previous work on the detection of magnetic helicity in astrophysical contexts has focused mainly on either magnetic fields of specific objects, such as the Sun \citep[see e.g.][and references therein]{zhang-2010} or astrophysical jets \citep[cf. e.g.][]{ensslin-2003, gabuzda-2004}, or cosmological primordial magnetic fields \citep[e.g.][]{kahniashvili-2005a, kahniashvili-2005b}. Two exceptions are the work by \citet{volegova-2010}, in which the use of Faraday rotation and synchrotron radiation for detecting magnetic helicity was suggested for the first time, and the work of \citet{kahniashvili-2006}, in which the use of charged ultra high energy cosmic rays of known sources is suggested for probing the three-dimensional structure of magnetic fields through which they pass. However, the sources of ultra high energy cosmic rays are not known yet and the applicability of this test is therefore limited.

The \textit{LITMUS} (\textbf{L}ocal \textbf{I}nference \textbf{T}est for \textbf{M}agnetic fields which \textbf{U}ncovers helice\textbf{S}) procedure for the detection of magnetic helicity suggested by \citet{junklewitz-2010} probes the local current helicity density $\vec{B}\cdot\vec{j}$, which for an ideally conducting plasma becomes
\begin{equation}
	\vec{B}\cdot\vec{j}\propto\vec{B}\cdot\left(\vec{\nabla}\times\vec{B}\right).
	\label{eq:helicity}
\end{equation}
Here, the magnetic field is denoted by $\vec{B}$ and the electric current density by $\vec{j}$. The test uses measurements of the Faraday depth and of the polarization direction of synchrotron radiation to probe the magnetic field components along the line of sight and perpendicular to it, respectively. Its simple geometrical motivation should make it applicable in a general setting, provided these quantities can be measured. The results depend only on the properties of the magnetic field along a line of sight and are therefore purely local in the two-dimensional sky projection. Our aim is to test this idea on observational as well as on simulated data, thereby determining the conditions under which the test will yield useful results.

This paper is organized as follows. In Sect.~\ref{sec:helicity-test}, the basic equations used in the \textit{LITMUS} test are reviewed. They are applied to observational data describing the galactic magnetic field in Sect.~\ref{sec:obs}, with special emphasis on a sophisticated reconstruction of the Faraday depth, described in Sect.~\ref{sec:reconstruction}. Section~\ref{sec:mock} is devoted to a thorough general assessment of the test's reliability. To this end it is applied to simulated observations of increasing complexity. Section~\ref{sec:planar} describes the application in a flat sky approximation, whereas Sect.~\ref{sec:spherical} examines all-sky simulations, finally arriving at complete simulations of the galactic setting in Sect.~\ref{sec:spherical-realistic}, where realistic electron distributions are added. We discuss our results and conclude in Sect.~\ref{sec:discussion}.

\section{The helicity test}
\label{sec:helicity-test}

For a thorough introduction into the ideas behind the \textit{LITMUS} test, the reader is referred to \citet{junklewitz-2010}. Here, we only summarize the resulting equations.

On the one hand side, synchrotron emission produced by cosmic ray electrons is used to probe the magnetic field component perpendicular to the line of sight. Its polarization is described by the complex field
\begin{equation}
	\label{Pdef}
	P=Q+iU=\left|P\right|e^{2i\chi},
\end{equation}
where $Q$ and $U$ are the usual Stokes parameters quantifying the linearly polarized components of the radiation with respect to some orthogonal coordinate system and $\chi$ is the polarization angle with respect to the first coordinate direction. On the other hand, the Faraday depth
\begin{equation}
	\label{phidef}
	\phi\propto\int_\textrm{LOS}n_\textrm{e}\vec{B}\cdot \rm{d}\vec{l}
\end{equation}
is used to probe the magnetic field component parallel to the line of sight (LOS).

A helical magnetic field will lead to a gradient of the Faraday depth that is parallel to the polarization direction of the synchrotron emission, as was argued in \citet{junklewitz-2010}. In order to compare the directions of the two quantities, this gradient is also formulated as a complex field
\begin{equation}
	\label{Gdef}
	G=\left(\left(\vec\nabla\phi\right)_x+i\left(\vec\nabla\phi\right)_y\right)^2=\left|G\right|e^{2i\alpha},
\end{equation}
with
\begin{equation}
	\alpha=\arctan\left(\frac{\left(\vec\nabla\phi\right)_y}{\left(\vec\nabla\phi\right)_x}\right),
\end{equation}
where the indices $x$ and $y$ denote its components with respect to the coordinates used. The helicity test that is performed in this work consists simply of multiplying $G$ with the complex conjugate of the polarization $P^*$. If the two angles $\chi$ and $\alpha$ differ by a multiple of $\pi$ (i.e. the gradient and the polarization direction are parallel), the product will be real and positive. If they differ by an odd multiple of $\pi/2$ (i.e. the two directions are perpendicular), it will be real and negative. Any orientation in between will produce varying real and imaginary parts in the product. Thus, observational directions along which a magnetic field is helical are indicated by a positive real part and a vanishing imaginary part of the product. Averaging over all observational directions will give an indication of the global helicity of the field.

It was furthermore shown by \citet{junklewitz-2010} that the ensemble average of this product over all magnetic field realizations given a magnetic correlation tensor (and therefore a helicity power spectrum) is a measure for the squared integrated spectral current helicity density
\begin{equation}
	\left<GP^*\right>_{\vec{B}}\propto\left(\int_0^\infty\textrm{d}k\frac{\epsilon_H(k)}{k}\right)^2,
\end{equation}
with large scales weighted more strongly than small scales.

\section{Application to galactic observations}
\label{sec:obs}

In this section, we try to answer the question whether the magnetic field of the Milky Way is helical by applying the \textit{LITMUS} test to the available observational data. Since the magnetic field is localized in a region that surrounds the observer, all relevant quantities will be given as fields on the sphere $\mathcal{S}^2$, i.e. as functions of the observational direction, specified by two angles $\vartheta$ and $\varphi$, which are taken to represent the standard spherical polar coordinates in a galactic coordinate system.

\subsection{Observational data}

For the synchrotron emission, we use the data gathered by the \textit{WMAP} satellite after seven years of observations\footnote{The data are available from NASA's Legacy Archive for Microwave Background Data Analysis at \texttt{http://lambda.gsfc.nasa.gov}.}, described in \citet{page-2007}. Since the foreground synchrotron emission is most intense at low frequencies, we use the measurement in the K-Band, which is centered at a frequency of $\nu=23\textrm{ GHz}$. Furthermore, we assume that the detected polarized intensity is solely due to galactic synchrotron emission. Thus, the Stokes $Q$ and $U$ parameter maps (defined with respect to the spherical polar coordinate directions $\hat{\vec{e}}_\vartheta$ and $\hat{\vec{e}}_\varphi$ in the galactic coordinate system) can be simply combined according to Eq.~\eqref{Pdef} to give the complex quantity $P$ whose argument is twice the rotation angle of the plane of polarization with respect to the $\hat{\vec{e}}_\vartheta$-direction
\begin{equation}
	\chi(\vartheta,\varphi)=\frac{1}{2}\arctan\left(\frac{\textrm{Im}(P(\vartheta,\varphi))}{\textrm{Re}(P(\vartheta,\varphi))}\right)
\end{equation}
\citep[cf.][]{junklewitz-2010}.

There are several depolarizing effects that have to be considered when dealing with polarization data. Faraday depolarization, which is important only at low frequencies due to the proportionality of the Faraday rotation angle to the square of the wavelength, can be safely neglected in the K-Band. Depolarization effects due to different magnetic field orientations along the line of sight are certainly present. However, they are present as well in the numerical test cases presented in Sect.~\ref{sec:mock}, which yield good results. Additionally, depolarization due to the finite beam-size of the \textit{WMAP} satellite and the finite pixel size of the polarization maps used in this study can play a role. The only way to limit this effect is to use higher resolution maps, ultimately necessitating the use of \textit{Planck} data in the future.

In order to construct a map of the Faraday depth, we use the catalog of rotation measurements provided by \citet{taylor-2009}\footnote{The catalog is available ~at \texttt{http://www.ucalgary.ca/ras/rmcatatlogue}.}. These provide an observational estimate of the Faraday depth for certain directions in the sky where polarized radio point-sources could be observed. Since the catalog encompasses a large number (37\,543) of point-sources, it paints a rather clear picture of the structure of the Faraday depth. However, Earth's shadow prevents observations in a considerably large region within the southern hemisphere.

\subsection{Reconstructing the Faraday depth map}
\label{sec:reconstruction}

The reconstruction is conducted according to the \textit{critical filter} method first presented in \citet{ensslin_frommert-2010}. A more elegant derivation of the same filter can be found in \citet{ensslin_weig-2010}. Since this formalism takes into account available information on the statistical properties of the signal in the form of the power spectrum, it is able to interpolate into regions where no direct information on the signal is provided by the data, such as the shadow of Earth in this case. Furthermore, it takes into account the available information on the uncertainty of the measurements. All in all it is expected to lead to a reconstructed map of the Faraday depth that is much closer to reality than e.g. a map in which the data were simply smoothed to cover the sphere. Small-scale features that are lost in such a smoothing process are for example reproduced by the \textit{critical filter} algorithm.

\begin{figure}
	\resizebox{\hsize}{!}{\input{profilea.tex}}
	\caption{Vertical galactic profile $p(\vartheta)$ of the Faraday depth.}
	\label{fig:profile}
\end{figure}
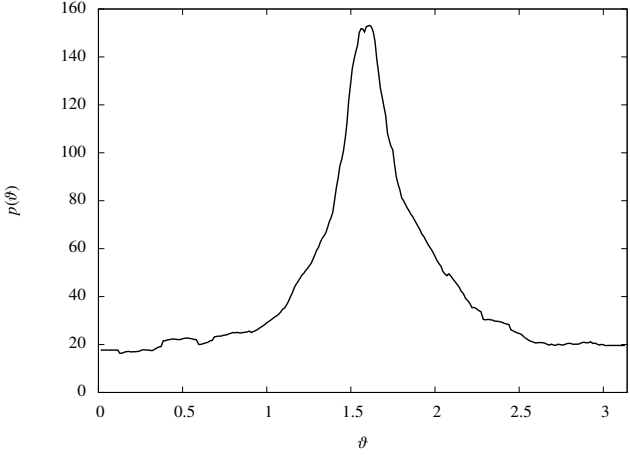

\subsubsection{Data model}

The field that is to be reconstructed here is the sky-map of the Faraday depth. In order to apply the \textit{critical filter} formula, the signal should be an isotropic Gaussian field. Since the Faraday depth clearly is larger along directions passing through the galactic plane, the condition of isotropy is not satisfied. Therefore, a vertical profile is calculated by binning the observations into intervals $\left[\vartheta_i,\vartheta_i+\Delta\vartheta\right)$, calculating the root mean square rotation measure value for each bin and smoothing the resulting values to obtain a smooth function $p(\vartheta)$. The result is shown in Fig.~\ref{fig:profile}. This profile is used to approximatively correct the anisotropies induced by the galactic structure and the resulting signal field
\begin{equation}
	\label{sdef}
	s(\vartheta,\varphi)=\frac{\phi(\vartheta,\varphi)}{p(\vartheta)}
\end{equation}
is assumed to be isotropic and Gaussian with a covariance matrix $S$. The Gaussian covariance matrix is determined solely by the angular power spectrum coefficients $C_l$, the reconstruction of which is part of the problem at hand.

The data $d$, i.e. the rotation measure values in the catalog, are taken to arise from the signal $s$ by multiplication with a response matrix $R$, which consists of a part encoding the specific directions in which the signal field is probed in order to produce the measurements and another part that is a simple multiplication with the vertical profile $p(\vartheta)$. Additionally, a Gaussian noise component $n$ is assumed with a covariance matrix $N=\textrm{diag}(\sigma_1^2,\sigma_2^2,\dots)$, where $\sigma_i$ is the one sigma error bar for the $i$th measurement in the catalog. Thus, the data are given by\footnote{The discretized version used in the implementations is $d_i=\sum_{j}R_{ij}s_j+n_i=\sum_{j}\tilde{R}_{ij}p_js_j+n_i$, where the index $j$ determines a pixel on the sphere, so that $s_j=s(\vartheta_j,\varphi_j)$ and $p_j=p(\vartheta_j)$.}
\begin{equation}
	d=Rs+n=\tilde{R}ps+n.
\end{equation}
Recent discussions in the literature \citep[see e.g.][]{stil-2010} have shown, however, that the error estimates as quoted in the catalog are probably too low. In addition, any contribution to the measured data from intrinsic Faraday rotation within the sources will further increase the error budget since the signal field in this context is only the contribution of the Milky Way. We therefore adapt the error bars of \citet{taylor-2009} according to the formula
\begin{equation}
	\sigma^{\textrm{(corrected)}}=\sqrt{\left(f_\sigma\sigma\right)^2+\left(\sigma^{\textrm{(int)}}\right)^2},
	\label{eq:sigma-correct}
\end{equation}
respectively. Here, the factor $f_\sigma$ accounts for the general underestimation of the errors in the catalog of \citet[]{taylor-2009}, whereas the additive constant $\sigma^{\textrm{(int)}}$ represents the average contribution of the sources' intrinsic Faraday rotation. As numerical values, we use $f_\sigma=1.22$, which was found by \citet{stil-2010} by comparing the data of \citet[]{taylor-2009} and \citet[]{mao-2010} on sources that are contained in both catalogs, and $\sigma^{\textrm{(int)}}=6.6\,\textrm{m}^{-2}$, which corresponds to the upper end of the numbers found by \citet{schnitzeler-2010}. Since the error contribution from the internal Faraday rotation is not correlated with the measurement error, the two contributions add up quadratically in the noise covariance matrix.

\subsubsection{Reconstruction method}

\begin{figure*}
	\centering
	\includegraphics[width=17cm]{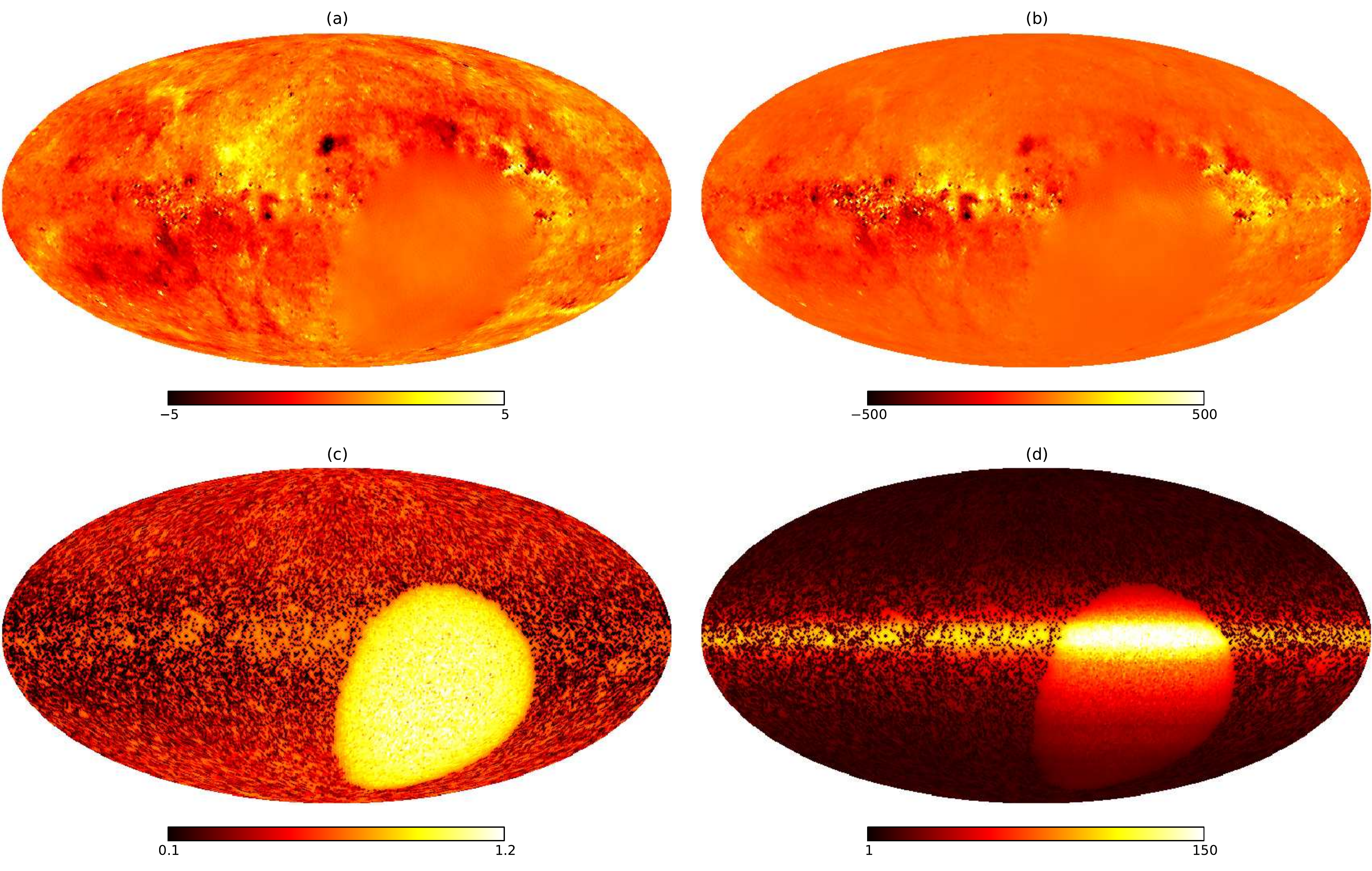}
	\caption[]{Results of the reconstruction of the signal field and Faraday depth. The left column shows the posterior mean of the signal field $m$ (panel (a)) and its one-sigma uncertainty $\sqrt{\hat{D}}$ (panel (c)). The right column shows the resulting map of the Faraday depth $pm$ (panel(b)) and the corresponding one-sigma uncertainty $\sqrt{p^2\hat{D}}$ (panel (d)) in $\textrm{m}^{-2}$.}
	\label{fig:maps}
\end{figure*}

For signal and noise fields with a Gaussian prior probability distribution, the posterior probability distribution of the signal field is again a Gaussian, i.e. it is of the form $\mathcal{P}(s|d,C_l)=\mathcal{G}(s-m,D)$, where a multivariate Gaussian probability distribution with covariance matrix $X$ is denoted by
\begin{equation}
	\mathcal{G}(x,X)=\frac{1}{\left|2\pi X\right|^{1/2}}\exp\left(-\frac{1}{2}x^\dagger X^{-1}x\right),
\end{equation}
where the $\dagger$ symbol denotes a transposed and complex conjugated quantity. In order to reconstruct the mean signal field $m=\left<s\right>$, where the brackets denote the posterior mean, it is also necessary to reconstruct the angular power spectrum $C_l$. To do so, the critical filter formulas
\begin{equation}
	m=Dj
	\label{eq:WF}
\end{equation}
and
\begin{equation}
	C_l=\frac{1}{2l+1}\textrm{tr}\left(\left(mm^\dagger+D\right)S_l\right)
	\label{eq:Cl}
\end{equation}
are iterated, starting with some initial guess for the power spectrum. Here, the signal covariance matrix is expanded as $S=\sum_lC_lS_l$, where $S_l$ is the projection onto the spherical harmonic components with index $l$. Furthermore, $D$ is the posterior covariance matrix,
\begin{equation}
	D=\left(S^{-1}+R^\dagger N^{-1}R\right)^{-1},
	\label{eq:Ddef}
\end{equation}
and $j$ is the information source term,
\begin{equation}
	j=R^\dagger N^{-1}d.
\end{equation}

Since the critical filter is on the brink of exhibiting a perception threshold \citep[cf.][]{ensslin_frommert-2010} and it is generally more desirable to overestimate a power spectrum entering a filter than to underestimate it, the coefficients $C_l$ are subjected to a procedure in which the value of $C_l$ is replaced by $\max\left\{C_{l-1},C_l,C_{l+1}\right\}$ after each iteration step. The advantage of overestimating the power spectrum can be seen by considering the limit of $C_l\rightarrow\infty$ in Eq.~\eqref{eq:WF} and~\eqref{eq:Ddef}. For high values of $C_l$, the first term in Eq.~\eqref{eq:Ddef} can be neglected and Eq.~\eqref{eq:WF} becomes
\begin{equation}
	m_{C_l\rightarrow\infty}=s+R^{-1}n.
\end{equation}
Thus, by overestimating the power spectrum the importance of its exact shape is diminished and the reconstruction will instead follow the information given directly by the data more closely. Considering an extreme underestimation of the power spectrum, $C_l\rightarrow0$, on the other hand, would lead to
\begin{equation}
	m_{C_l\rightarrow0}=0,
\end{equation}
suppressing the information given by the data.

\subsubsection{Calculating $\left<G\right>_{\mathcal{G}(s-m,D)}$}
\label{sec:calc-G}

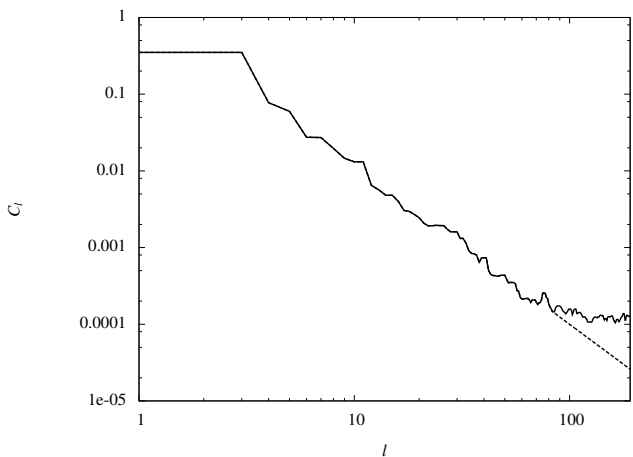
\begin{figure}
	\resizebox{\hsize}{!}{\input{Cla.tex}}
	\caption{Angular power spectrum of the signal field $s$. The solid curve shows the resulting power spectrum as calculated with the \textit{critical filter} formalism. The dashed line depicts the signal power spectrum used to generate large-scale uncertainty corrections for the calculation of $G$ according to Sect.~\ref{sec:calc-G}.}
	\label{fig:Cl}
\end{figure}

\begin{figure*}
	\centering
	\includegraphics[width=17cm]{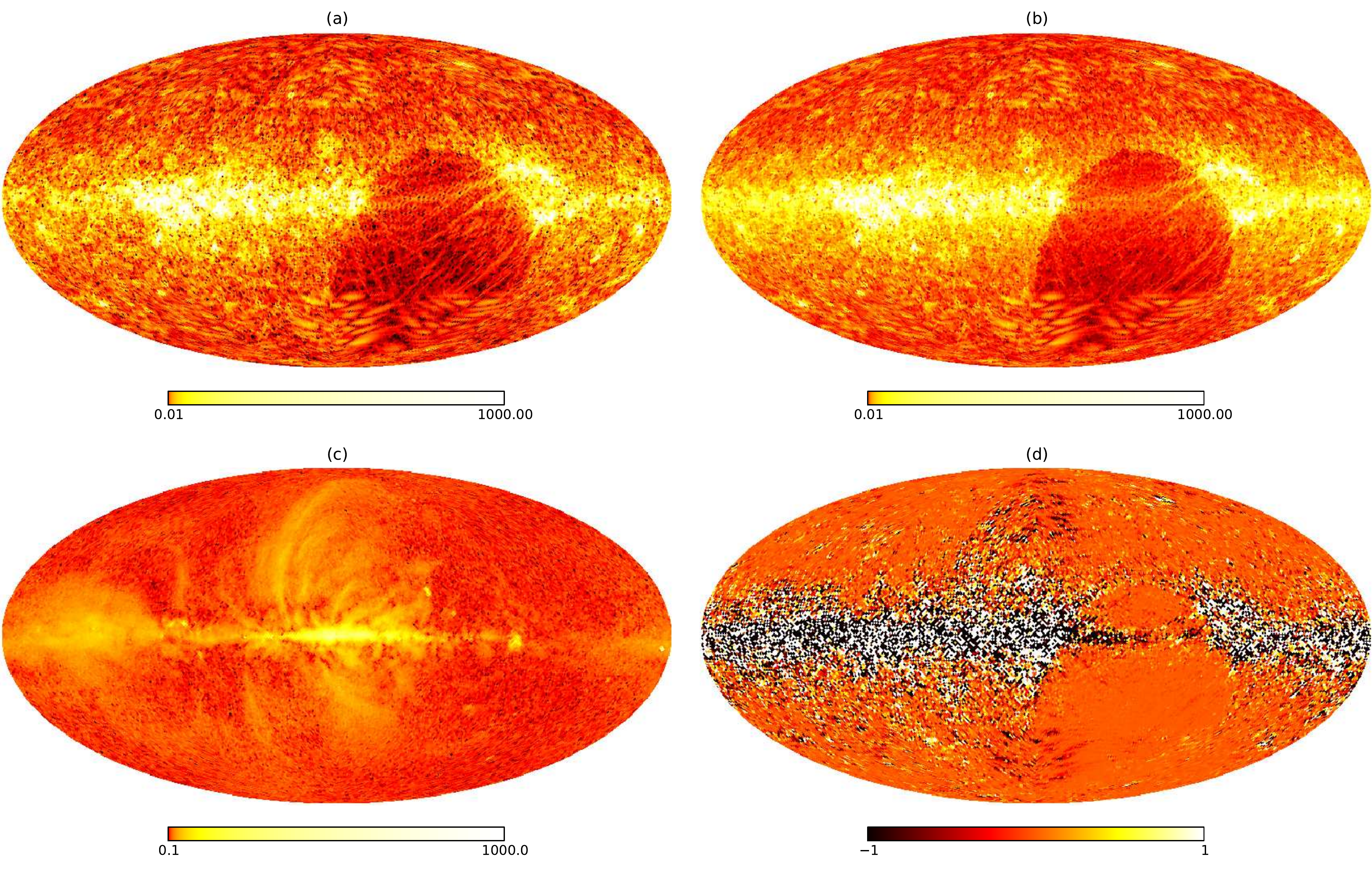}
	\caption[]{Results of the calculation of the gradient field $G$ and the application of the \textit{LITMUS} test. The first row shows estimates for the ablolute values of the field $G$. Panel (a) shows $G$ as calculated directly from the posterior mean $m$ of the signal field, whereas panel (b) takes into account large-scale uncertainty corrections as described in Sect.~\ref{sec:calc-G}. Panel (c) shows the absolute value of the polarized intensity $P$ as measured by the \textit{WMAP} satellite and panel (d) shows the result of the \textit{LITMUS} test, namely $\mathrm{Re}\left(GP^*\right)$, in arbitrary units, using $G$ and $P$ as shown in panels (b) and (c), respectively. Note the logarithmic color code of panels (a)--(c).}
	\label{fig:litmus-res}
\end{figure*}

Once the posterior mean of the signal is reconstructed, the corresponding field $G$ can be calculated according to Eq.~\eqref{Gdef}, assuming $\phi=m$. However, since Eq.~\eqref{Gdef} is a nonlinear transformation of the Faraday depth, that is not the same as the posterior mean of $G$. This can be estimated by averaging over a large number of samples, according to
\begin{equation}
	\left<G(s)\right>_{\mathcal{G}(s-m,D)}\approx\frac{1}{N_\textnormal{samples}}\sum_{k=1}^{N_\textnormal{samples}}G(s_k).
        \label{eq:sampling}
\end{equation}
Here, the samples $s_k$ are drawn from the posterior probability distribution $\mathcal{G}(s-m,D)$. In our implementation we use the method of \citet[]{jasche-2010} to generate these samples, i.e. we generate corrections $y_k$ to the posterior mean $m$ by drawing signal and noise realizations from the respective prior probability distributions,
\begin{equation}
	\tilde{s}_k\hookleftarrow\mathcal{P}(s|S),
	\label{eq:draw-signal}
\end{equation}
\begin{equation}
	n_k\hookleftarrow\mathcal{P}(n|N),
\end{equation}
combining them into a data realization according to
\begin{equation}
	d_k=R\tilde{s}_k+n_k,
\end{equation}
and calculating the difference between the signal realization and the Wiener filtered data realization
\begin{equation}
	y_k=\tilde{s}_k-DR^\dagger N^{-1}d_k.
\end{equation}
The signal realizations
\begin{equation}
	s_k=m+y_k
\end{equation}
are shown by \citet[]{jasche-2010} to follow the posterior probability distribution. For reasons of numerical feasibility, we consider only large scale corrections to the posterior mean in our calculation, i.e. we introduce a small-scale power cut-off for the signal power spectrum that describes the prior probability distribution used in Eq.~\eqref{eq:draw-signal}.

Thus, given $m$ and $D$, the posterior mean for the Faraday depth is given by $\left<\phi\right>=pm$, its one sigma error bars by $\pm p\sqrt{\hat{D}}$, and the posterior mean for the field $G$ by the above sampling approximation, where the signal samples $s_k$ are multiplied with the galactic profile function $p(\theta)$ to give realizations of the Farady depth. Here, $\hat{D}$ is the vector that contains the diagonal elements of the matrix $D$.

\subsection{Results}
\label{sec:obs-results}

Figures~\ref{fig:maps}~--~\ref{fig:litmus-res} summarize the results of the Faraday depth reconstruction and the application of the \textit{LITMUS} test to these data. All calculations are conducted at a \textsc{HEALPix}\footnote{The \textsc{HEALPix} package is available from \texttt{http://healpix.jpl.nasa.gov}.} resolution $N_{\textrm{side}}=64$. The left column of Fig.~\ref{fig:maps} shows the reconstructed signal field as well as its uncertainty as they are calculated using the \textit{critical filter} method. Evidently, the reconstruction method is able to extrapolate from the available information into regions where no data are taken, i.e. the Earth's shadow in the lower right of the projection. However, only structures on scales comparable to the extent of the region without information are reconstructed within it and the reconstruction's uncertainty becomes large in this region, as well as near the galactic poles, where the signal-response to noise ratio of the data is low.

The outcome for the angular power spectrum is shown in Fig.~\ref{fig:Cl}. Clearly, the \textit{critical filter} predicts a large amount of power on small scales. Therefore the rather high values seen in Fig.~\ref{fig:maps}.c are mainly due to uncertainty on small scales. This may in part be due to an underestimation of the error bars of extreme data points which are not accounted for by our rather crude correction of the error bars as described by Eq.~\eqref{eq:sigma-correct}.\footnote{A thorough study of the problem of reconstructing a signal field with unknown power spectrum from data with unknown error bars will be presented in a future paper.}

The right column of Fig.~\ref{fig:maps} shows the reconstructed Faraday depth and its uncertainty, obtained from the left column of Fig.~\ref{fig:maps} by a simple multiplication with the galactic profile function $p(\vartheta)$. Clearly, the field $\left<\phi\right>$ takes on only low values within the information-less region, whereas, again, its uncertainty is especially large there.

Figure~\ref{fig:litmus-res} shows the two fields used in the \textit{LITMUS} test as well as its result. The field $G$ calculated according to the sampling prescription, Eq.~\eqref{eq:sampling}, is shown in panel (b). In the sampling procedure, large-scale uncertainty corrections that are described by the signal power spectrum shown in Fig.~\ref{fig:Cl} are taken into account. For comparison, the field $G$ as calculated from the posterior mean of the signal field without any uncertainty corrections is shown in panel (a). The main effect of the uncertainty corrections is to increase the values within the region where no data points were measured and in the polar regions. The spatial structure of $G$, however, remains largely unchanged.

\begin{figure}
	\resizebox{\hsize}{!}{\input{rotation-obsa.tex}}
	\caption{Rotational curve for $\mathrm{Re}\left<GP^*\right>_{\mathcal{S}^2}$ in arbitrary units.}
	\label{fig:rotation-obs}
\end{figure}
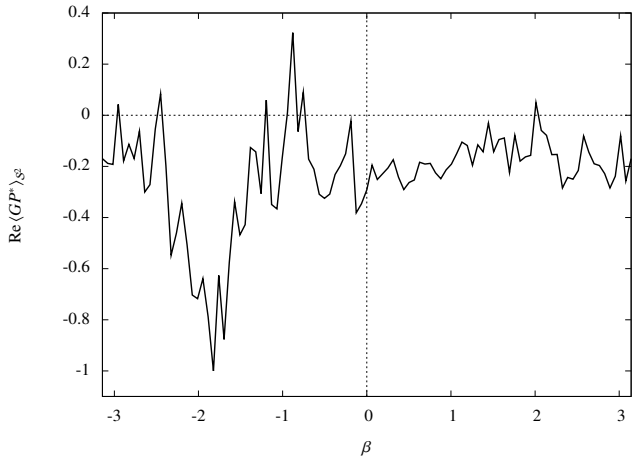

Finally, the real part of the product $GP^*$ is shown in the last panel of Fig.~\ref{fig:litmus-res}. By visual inspection, it is not immediately clear whether positive values dominate this image. Taking the spatial average $\mathrm{Re}\left<GP^*\right>_{\mathcal{S}^2}$ does in fact yield a negative value. Nevertheless, we performed another test for helicity. In order to assess helicity on small scales, the two fields $G$ and $P$ are rotated with respect to one another about an angle $\beta$ around the galactic axis and their product is again averaged over the whole sky. For a helical magnetic field, this procedure should result in a curve $\mathrm{Re}\left<GP^*\right>_{\mathcal{S}^2}(\beta)$ that exhibits a maximum at $\beta=0$, since the correlations between the fields $G$ and $P$ are expected to be local. The resulting curve for this test is shown in Fig.~\ref{fig:rotation-obs}. Not only does it not take on its global maximum at $\beta=0$, but the point $\beta=0$ does not seem to be special in any way. Therefore, no indication of helicity is found.

We also performed the analysis with a different function $p(\vartheta)$, created by doubling the smoothing length in its calculation. The results of the reconstructed map of the Faraday depth changed only slightly, i.e. the variance of the difference between the reconstructions with the two different profiles is $0.65\%$ of the variance of the original $\left<\phi\right>$-map shown in Fig.~\ref{fig:maps}.b.

\section{Application to simulated data}
\label{sec:mock}

In order to check whether the nondetection of helicity in the previous section allows the conclusion that the galactic magnetic field is in fact nonhelical, we now apply the same helicity test to a number of artificially generated magnetic fields with known helicity.

\subsection{Planar implementation}
\label{sec:planar}

The most simple setting that can be considered is the observation of a well localized magnetic field structure. In the limit of great distances between the magnetic field under consideration and the observer, the lines of sight penetrating the field become parallel. We assume the field to be contained in a cubic box which is oriented along the lines of sight.

The field in the box is generated by the \textsc{garfields} code \citep[first applied in][]{kitaura-2008}. This code draws the three cartesian components of the magnetic field in Fourier space independently of a common power spectrum, assumed here to be a Kolmogorov-type spectrum of the form $P_B(k)\propto k^{-5/3-2}$, according to Gaussian statistics. In order to produce a magnetic field without divergence, its frequency components parallel to the respective $k$-vector are then subtracted
\begin{equation}
	\vec{B}_\textrm{div-free}(\vec{k})\propto\vec{B}(\vec{k})-\vec{k}\frac{\vec{k}\cdot\vec{B}(\vec{k})}{k^2}.
\end{equation}
A degree of helicity is then imprinted onto the field by applying the formula
\begin{equation}
	\vec{B}_\textrm{div-free,hel}(\vec{k})\propto\vec{B}_\textrm{div-free}(\vec{k})+\eta\frac{i\vec{k}\times\vec{B}_\textrm{div-free}(\vec{k})}{k},
\end{equation}
where $\eta=0$ leaves the field unaffected and $\eta=\pm1$ produces the highest degree of helicity.

Finally, we assume the thermal and cosmic ray electron densities to be constant throughout the box. Thus, the observables $Q$, $U$, and $\phi$ can be obtained by simply integrating the appropriate magnetic field components along the box direction associated with the line of sight. Then the complex quantities $G$ and $P$ are easily calculated and multiplied, yielding two-dimensional images of $GP^*$. This procedure is conducted for various realizations of random magnetic fields both without helicity ($\eta=0$) and with maximal helicity ($\eta=1$).

\subsubsection{Results}
\label{sec:planar-results}

\begin{figure*}
	\centering
	\includegraphics[width=17cm]{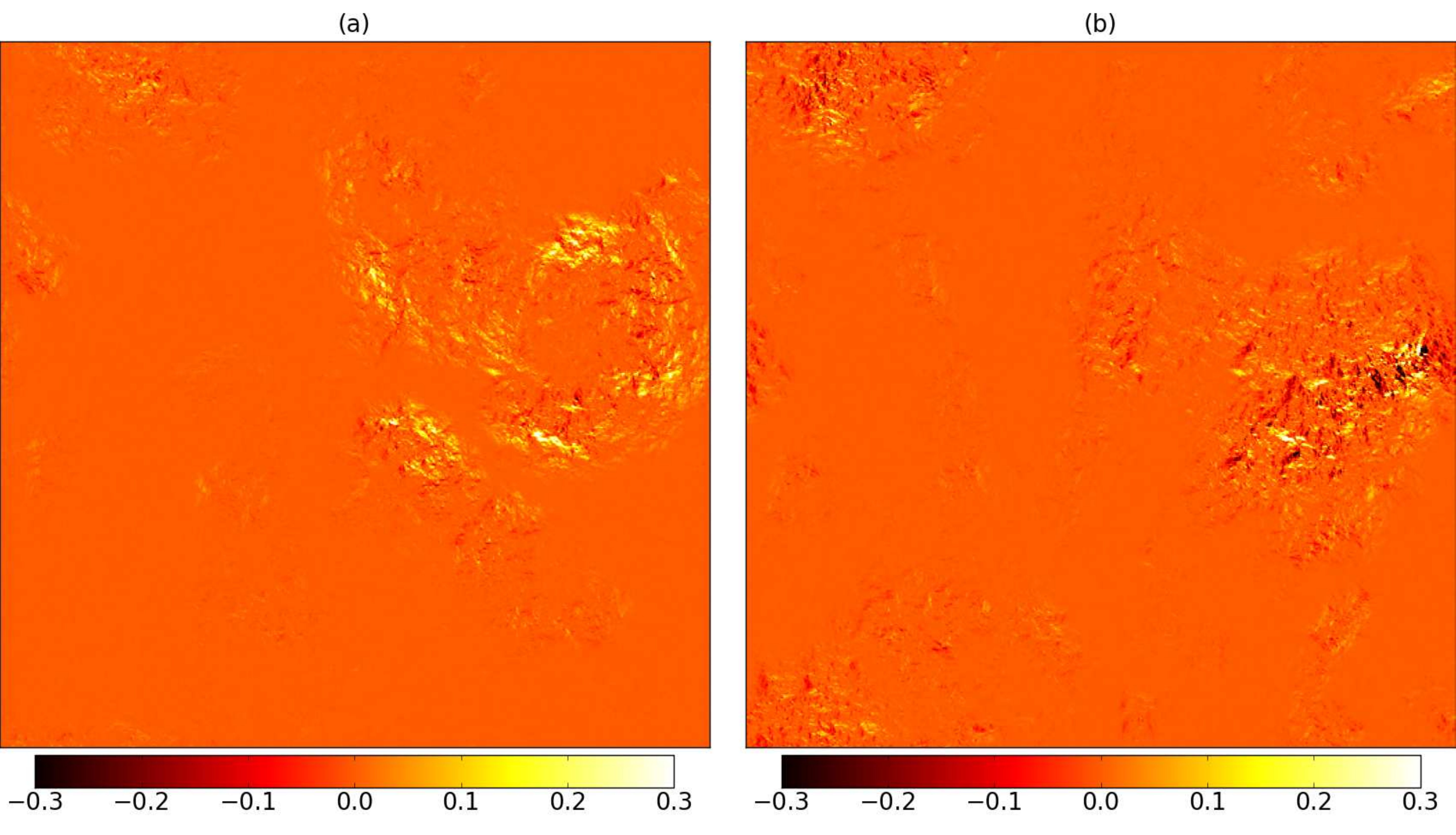}
	\caption{Maps of $\mathrm{Re}\left(GP^*\right)$ for a particular magnetic field realization in arbitrary units. Panel (a) shows the case with maximal helicity ($\eta=1$), panel (b) the one without helicity ($\eta=0$).}
	\label{fig:planar}
\end{figure*}

The resulting images for one random magnetic field realization are shown in Fig.~\ref{fig:planar}. The cube was discretized for the calculation into $512^3$ pixels. It can already be seen by eye that positive values of $\mathrm{Re}\left(GP^*\right)$ dominate in the case with maximal helicity (panel (a)), whereas in the case without helicity (panel (b)), positive and negative values seem to be roughly equally represented. Calculating the spatial averages over the whole square yields $\mathrm{Re}\left<GP^*\right>_\square=2.3\cdot 10^{-3}$ and $\mathrm{Re}\left<GP^*\right>_\square=-1.5\cdot 10^{-4}$ for the case with and without helicity, respectively.

We calculated this spatial average for the results of the \textit{LITMUS} test applied to 100 different random magnetic field realizations, both with and without helicity. Averaging these values for the helical fields and for the nonhelical fields separately yields a positive value in the helical case. Normalizing all values such that this average becomes equal to one yields
\[\left<\mathrm{Re}\left<GP^*\right>_{\square}\right>_\mathrm{samples}=1.0,~~\sigma_{\mathrm{Re}\left<GP^*\right>_{\square}}=0.63\] in the case with helicity and
\[\left<\mathrm{Re}\left<GP^*\right>_{\square}\right>_\mathrm{samples}=-0.025,~~\sigma_{\mathrm{Re}\left<GP^*\right>_{\square}}=0.34\] in the case without. Clearly, the \textit{LITMUS} test yields positive results if applied to helical fields, whereas its results fluctuate around zero if applied to nonhelical fields. This is exactly the behavior that should be expected and the basic functioning of the \textit{LITMUS} test is thereby demonstrated in this setting.

\subsection{Spherical implementation}
\label{sec:spherical}

As a next step, the applicability of the \textit{LITMUS} test is checked for magnetic fields surrounding the observer, as in the case of the galactic field. Again, several sets of mock observations are produced. These simulations are conducted using the \textsc{hammurabi} code \citep[see][]{waelkens-2009} in connection with the \textsc{garfields} code\footnote{Both codes are available from \texttt{http://www.mpa-garching.mpg.de/hammurabi/hammurabi11}.}. The \textsc{hammurabi} code allows for a large scale analytic field model and an additional Gaussian random field component, which can be generated by the \textsc{garfields} code with a preset degree of helicity as described in Sect.~\ref{sec:planar}. It then integrates the different field components, weighted with the appropriate electron density, along radial lines of sight and produces sky maps of simulated observations of the Stokes parameters $Q$ and $U$ and Faraday depth $\phi$ (among others), thus providing all necessary ingredients to perform the \textit{LITMUS} procedure.

\subsubsection{Constant electron densities}
\label{sec:spherical-constant}

For simplicity, we start again by setting the densities of the thermal electrons and the cosmic ray electrons to constant values throughout the simulated galaxy.

\paragraph{Gaussian random field.}
\label{sec:spherical-constant-random}

First, in order to apply the test to a field with a well-defined degree of helicity, the field strength of the large scale analytic component is set to zero, such that the simulated galactic field is a purely random one with a chosen degree of helicity. As in the planar case, we choose either no helicity ($\eta=0$) or maximal helicity ($\eta=1$).

\begin{figure*}
	\centering
	\includegraphics[width=17cm]{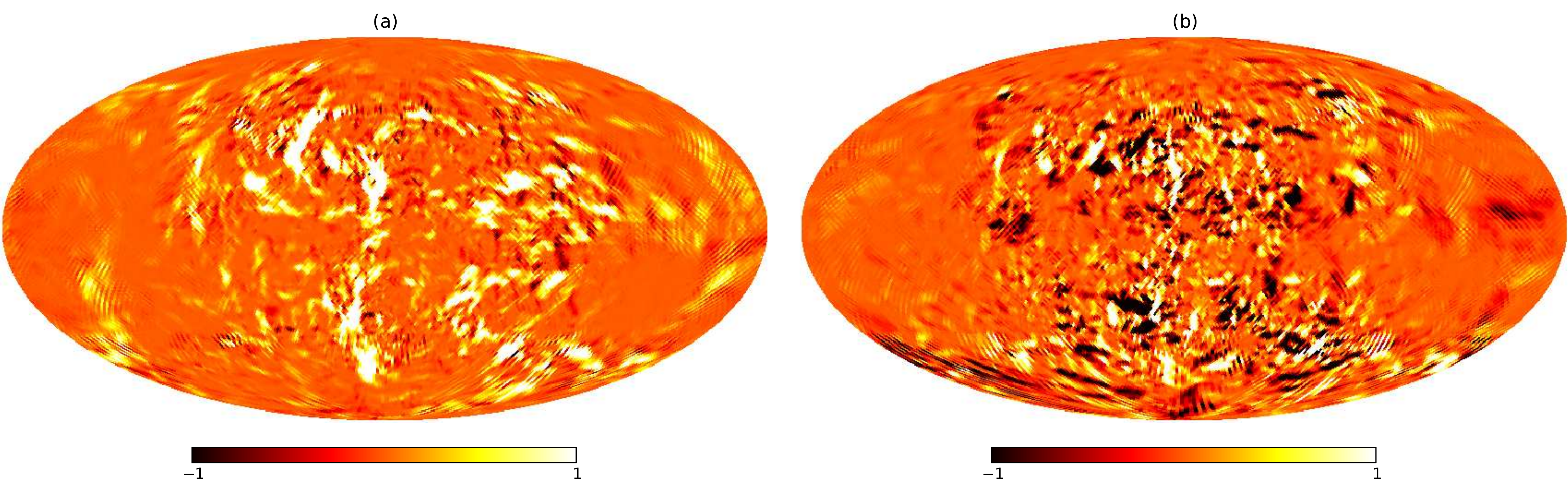}
	\caption{Maps of $\mathrm{Re}\left(GP^*\right)$ for a particular magnetic field realization in arbitrary units in a spherical setting. Panel (a) shows the case with maximal helicity ($\eta=1$), panel (b) the one without helicity ($\eta=0$).}
	\label{fig:spherical}
\end{figure*}

Figure~\ref{fig:spherical} shows the maps of $\mathrm{Re}\left(GP^*\right)$ for one particular Gaussian random magnetic field with a power law index of $-5/3$, with helicity and without helicity, respectively. As for the planar implementation, one can immediately see that positive values dominate in the helical case and positive and negative values are roughly equally represented in the nonhelical case.

Figure~\ref{fig:rotation-hel-nohel-noel} shows the results of the rotational test described in Sect.~\ref{sec:obs-results} for this particular magnetic field realization. Clearly, the spatial average $\mathrm{Re}\left<GP^*\right>_{\mathcal{S}^2}$ takes on a sharp maximum at $\beta=0$ and is positive in the case with helicity, while it does not have a maximum there and in fact happens to be negative in the case without helicity. This is exactly the result expected from the \textit{LITMUS} test.

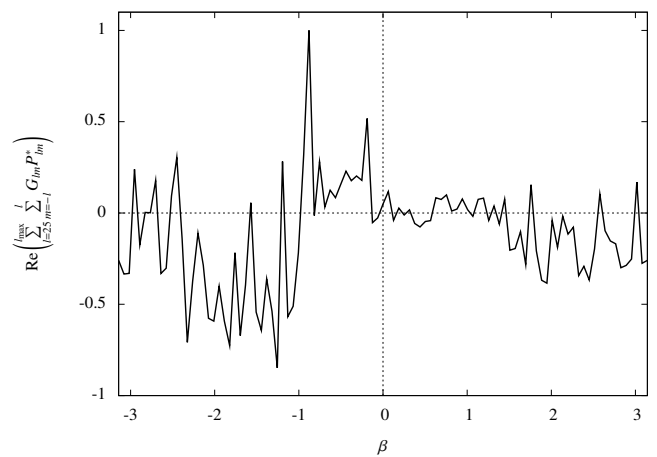
\begin{figure}
	\resizebox{\hsize}{!}{\input{rotation-obs-small-scalea.tex}}
	\caption{Same as Fig.~\ref{fig:rotation-obs}, only with large-scale contributions neglected.}
	\label{fig:rotation-obs-small-scale}
\end{figure}

The sharpness of the peak in Fig.~\ref{fig:rotation-hel-nohel-noel} indicates that the helicity is to be found in small-scale features. As a test of this assertion, we calculated the spherical multipole components $G_{lm}$ and $P_{lm}$ of the complex gradient and polarization fields. Note that the spatial average over the product of the fields is proportional to the sum of the products of the multipole components, i.e.
\begin{equation}
	\textrm{Re}\left<GP^*\right>_{\mathcal{S}^2}=\frac{1}{4\pi}\textrm{Re}\left(\sum\limits_{l=0}^{l_\mathrm{max}}\sum\limits_{m=-l}^{l}G_{lm}P_{lm}^*\right),
	\label{eq:sum}
\end{equation}
where $l_\mathrm{max}$ is determined by the finite resolution of the map. If we now neglect the first terms in the sum, i.e. the small-$l$ contributions, we arrive at a spatial average over the product in which all large-scale features were neglected. The resulting rotational curves for the same magnetic fields used for Fig.~\ref{fig:rotation-hel-nohel-noel} are shown in Fig.~\ref{fig:rotation-hel-nohel-noel-small-scale}, where only multipole moments with $l\geq25$ were considered. It can be seen that this procedure further sharpens the peak at $\beta=0$ and strengthens it relative to other local maxima in the curves, thus facilitating the detection of small-scale helicity. The same result for the observational data studied in Sect.~\ref{sec:obs} is shown in Fig.~\ref{fig:rotation-obs-small-scale}. Clearly, there is still no sign of helicity in this case.

Furthermore, we created a set of 100 different Gaussian random magnetic fields, performed the \textit{LITMUS} test, and calculated the spatial average $\mathrm{Re}\left<GP^*\right>_{\mathcal{S}^2}$ for all of them. Each field realization was considered in a version without helicity and a version with maximal helicity, just as in the case of the planar implementation. Averaging over the 100 samples yields again a positive value in the helical case. Normalizing all values such that this average is equal to one yields \[\left<\mathrm{Re}\left<GP^*\right>_{\mathcal{S}^2}\right>_\mathrm{samples}=1.0,~~\sigma_{\mathrm{Re}\left<GP^*\right>_{\mathcal{S}^2}}=0.25\] in the helical case and \[\left<\mathrm{Re}\left<GP^*\right>_{\mathcal{S}^2}\right>_\mathrm{samples}=-0.27,~~\sigma_{\mathrm{Re}\left<GP^*\right>_{\mathcal{S}^2}}=0.23\] in the nonhelical case. This clearly underlines the success of the \textit{LITMUS} test in the spherical setting.

Since the magnetic field of the Milky Way is known to be much stronger in the vicinity of the galactic plane than in the halo region, we repeated the analysis with random realizations of Gaussian magnetic fields that are confined to the region within a vertical distance of $8~\mathrm{kpc}$ of the galactic plane. The results of the \textit{LITMUS} procedure for these fields still turn out to be a reliable indicator for magnetic helicity.

\begin{figure}
	\resizebox{\hsize}{!}{\input{rotation-hel-nohel-noela.tex}}
	\caption{Rotational curve for $\mathrm{Re}\left<GP^*\right>_{\mathcal{S}^2}$ in arbitrary units. The solid and dashed lines depict the results for the same magnetic field with helicity parameter $\eta=1$ and $\eta=0$ respectively. Constant electron densities are assumed.}
	\label{fig:rotation-hel-nohel-noel}
\end{figure}
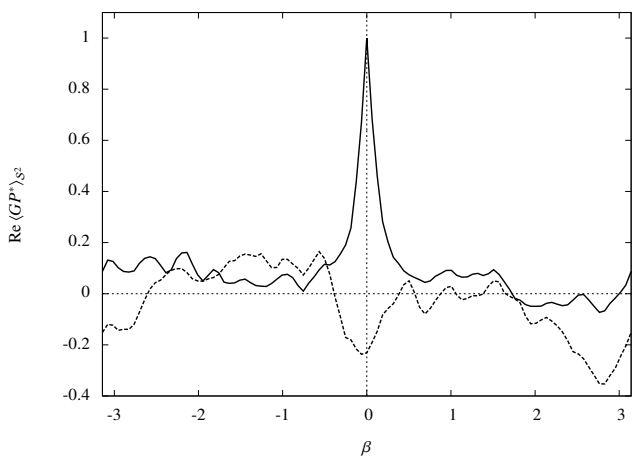

\begin{figure}
	\resizebox{\hsize}{!}{\input{rotation-hel-nohel-noel-small-scalea.tex}}
	\caption{Same as Fig.~\ref{fig:rotation-hel-nohel-noel}, only with large-scale contributions neglected.}
	\label{fig:rotation-hel-nohel-noel-small-scale}
\end{figure}
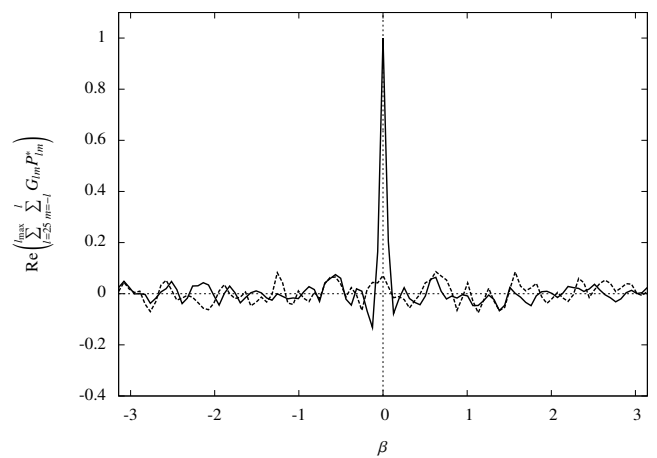

\paragraph{Large scale field models.}
\label{sec:spherical-constant-large}

As a next step, the (helical or nonhelical) random magnetic field component is switched off completely and replaced by an analytic large scale magnetic field model. Several sets of simulations are performed using different models for the galactic large scale field. For these analytic models, the current helicity can be calculated directly, giving an expectation as to whether the helicity test should produce positive results or not.

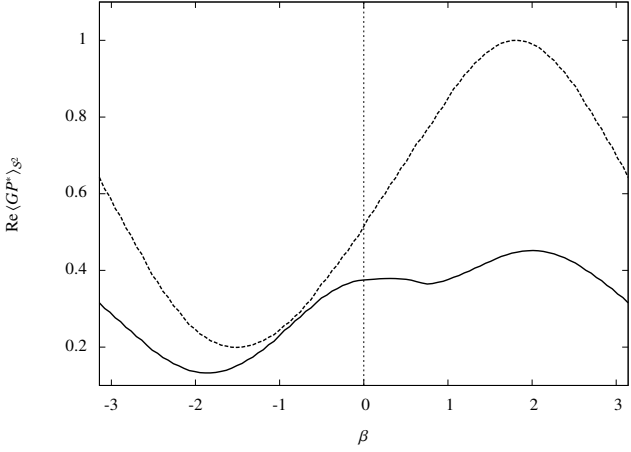
\begin{figure}
	\resizebox{\hsize}{!}{\input{rotation-WMAP-noela.tex}}
	\caption{Rotational curve for $\mathrm{Re}\left<GP^*\right>_{\mathcal{S}^2}$ in arbitrary units. The solid line corresponds to the best-fit magnetic field model of \citet{page-2007}, the dashed line to a model where the parameter $\chi_0$ is set to zero. Constant electron densities are assumed.}
	\label{fig:rotation-WMAP-noel}
\end{figure}

The results of the rotational helicity test are shown in Fig.~\ref{fig:rotation-WMAP-noel} for the large scale magnetic field model described in \citet{page-2007}, i.e.
\begin{equation}
	\begin{split}
	\vec{B}(r,\varphi,z)=B_0\left[\right.&\cos\left(\psi(r)\right)\cos\left(\chi(z)\right)\hat{\vec{e}}_r\\
		&+\sin\left(\psi(r)\right)\cos\left(\chi(z)\right)\hat{\vec{e}}_\varphi\\
		&\left.+\sin\left(\chi(z)\right)\hat{\vec{e}}_z\right]
	\end{split}
\end{equation}
in galactic cylindrical coordinates, where
\begin{equation}
	\psi(r)=\psi_0+\psi_1\ln\left(\frac{r}{8\mathrm{ kpc}}\right)
\end{equation}
and
\begin{equation}
	\chi(z)=\chi_0\tanh\left(\frac{z}{1\mathrm{ kpc}}\right).
\end{equation}
The solid curve corresponds to the parameters favored by \citet{page-2007}, namely $\chi_0=25^{\circ}$, $\psi_0=27^{\circ}$, and $\psi_1=0.9^{\circ}$. This corresponds to a simple flat spiral in the galactic plane which becomes more and more screw-like with vertical distance $z$ from the galactic plane, so that a slight degree of helicity is inherent in the field geometry. This can be verified by direct calculation according to Eq.~\eqref{eq:helicity}, yielding
\begin{equation}
	\vec{j}\cdot\vec{B}=B_0^2\frac{\sin\left(\chi(z)\right)\cos\left(\chi(z)\right)}{r}\left(\sin\left(\psi(r)\right)+\psi_1\cos\left(\psi(r)\right)\right),
	\label{eq:current-helicity-WMAP}
\end{equation}
which is nonzero for any generic point away from the galactic plane. The resulting line in Fig.~\ref{fig:rotation-WMAP-noel} is \em{not} \em a clear indication for this helicity. However, the curve is nevertheless sensitive to the angle $\chi_0$, which produces the helicity. Lowering its value, i.e. making the spirals more and more parallel to the galactic plane, changes the results of the \textit{LITMUS} test. The extreme case of $\chi_0=0$, i.e. $B_z=0$ everywhere, for which the value of Eq.~\eqref{eq:current-helicity-WMAP} becomes zero everywhere, is also shown in Fig.~\ref{fig:rotation-WMAP-noel}. This curve's value at $\beta=0$ is even more distinct from its maximum than in the case of the solid line. This example shows that while the results react in a systematic way on changes in the parameters, the test is not suited to detect helicity on the largest scales.

We used the model of \citet{page-2007} in the demonstration of this effect mainly because of its mathematical simplicity. More sophisticated models can be found e.g. in \citet{sun-2007}, \citet{jansson-2009}, \citet{jaffe-2010}, and references therein.

\subsubsection{The role of the electron densities}
\label{sec:spherical-realistic}

In order to get closer to a realistic model of the Milky Way, as a next step we replace the constant electron densities with realistic models. The \textsc{hammurabi} code allows the use of the NE2001 model for the thermal electron density \citep[cf.][]{cordes-2002,cordes-2003} to compute the Faraday depth and several analytic models for the cosmic ray electron density to compute the synchrotron emissivity \citep[see also][]{waelkens-2009}. In the calculations performed to obtain the results presented here, the cosmic ray electron distribution model of \citet{page-2007} was used.

\paragraph{Gaussian random field.}
\begin{figure}
	\resizebox{\hsize}{!}{\input{rotation-hel-nohela.tex}}
	\caption{Same as Fig.~\ref{fig:rotation-hel-nohel-noel}, only with realistic electron densities used in the calculation.}
	\label{fig:rotation-hel-nohel}
\end{figure}
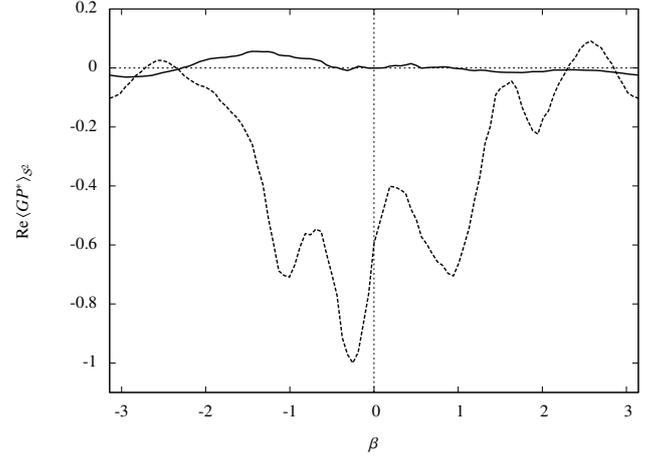

The resulting rotational curves, as calculated with the realistic electron distributions, for the case of the Gaussian random field are shown in Fig.~\ref{fig:rotation-hel-nohel}. Neither in the case with helicity ($\eta=1$), nor in the case without helicity ($\eta=0$) does the curve take on its maximum at $\beta=0$. This is the expected result in the latter case but contradicts the expectation in the former one. Therefore, the helicity imprinted onto the small scale magnetic field clearly fails to be detected by the test applied.

\paragraph{Large scale field models.}

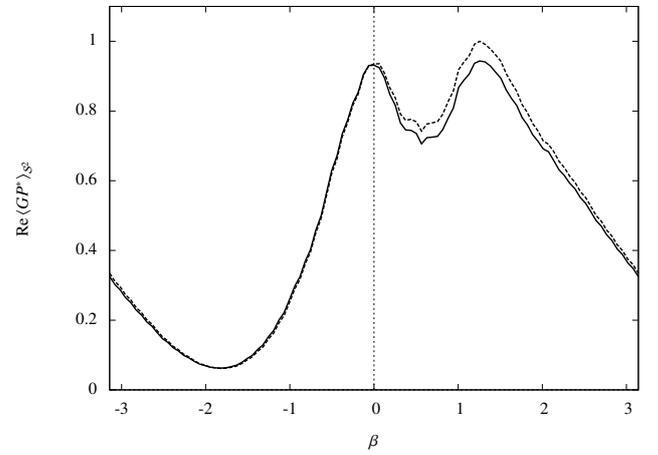
\begin{figure}
	\resizebox{\hsize}{!}{\input{rotation-WMAPa.tex}}
	\caption{Same as Fig.~\ref{fig:rotation-WMAP-noel}, only with realistic electron densities used in the calculation.}
	\label{fig:rotation-WMAP}
\end{figure}

The result for the large scale magnetic fields is shown in Fig.~\ref{fig:rotation-WMAP}. There actually seems to be a maximum in the vicinity of $\beta=0$ now. This is true, however, for the case of the planar spiral model ($\chi_0=0$) as well as for the model with helicity ($\chi_0=25^\circ$). Therefore, the proposed helicity test might under certain circumstances even indicate helicity on large scales where there is none, if the observer is surrounded by the field. Other magnetic field models with planar spirals, such as the bisymmetric (i.e. $\vec{B}(r,\varphi,z)=-\vec{B}(r,\varphi+\pi,z)$) spiral model of \citet{stanev-1997} lead to similar results.

\section{Discussion and conclusion}
\label{sec:discussion}

The present work presents the first application of the \textit{LITMUS} test for magnetic helicity proposed by \citet{junklewitz-2010} to actual data. The application of the test involves thoroughly reconstructing a map of the Faraday depth distribution, calculating its transformed gradient field $G$, creating a map of $GP^*$, averaging over this map, shifting the two fields with respect to each other to see whether any signal vanishes, and filtering out large-scale contributions for a better detection of small-scale helicity. This procedure, applied to observations of the Faraday depth and polarization properties of the synchrotron radiation within our own galaxy in Sect.~\ref{sec:obs}, does not show any signs of helicity in the Milky Way's magnetic field.

In order to assess the significance of this, the applicability of the test was probed in different artificial settings. The complexity of these settings was increased bit by bit to find out under what circumstances exactly the \textit{LITMUS} test yields reliable results. It was found that meaningful results can be achieved if the electron densities do not vary on the scales of the magnetic field, both in the regime of magnetic field structures whose distance from the observer is much greater than their extension, as shown in Sect.~\ref{sec:planar}, and in the regime of magnetic fields surrounding the observer, as shown in Sect.~\ref{sec:spherical-constant-random}. We showed that the performance of the \textit{LITMUS} test with regard to small-scale helicity is further improved by dropping the first few terms in Eq.~\eqref{eq:sum}. However, indications of helicity on large scales are unreliable, as shown in Sect.~\ref{sec:spherical-constant-large}. Furthermore, it was demonstrated in Sect.~\ref{sec:spherical-realistic} that any non-trivial electron density may distort the outcome of the test to a point where even small-scale helical structures fail to be detected. This is not too surprising since e.g. a variation in the thermal electron density will introduce a gradient in the Faraday depth that is not caused by the magnetic field structure. Therefore the nondetection of helicity for the galactic magnetic field does not necessarily mean that the field is nonhelical on small scales. It may be the case that small-scale fluctuations of the electron density introduce effects in the observational data that prevent the detection of helicity.

So, as a natural next step, the hunt for helicity in astrophysical magnetic fields should focus on a region that is small and/or homogeneus enough for the assumption of constant electron densities to hold at least approximatively. Although our work has shown that the helicity test that we studied is not suitable for all astrophysical settings, we are confident that it may nevertheless yield useful results if applied in a setting with constant electron densities.

As a side effect of this paper, it was demonstrated in Sect.~\ref{sec:reconstruction} that the method proposed by \citet{ensslin_frommert-2010} to reconstruct a Gaussian signal with unknown power spectrum is very well suited for practical application.

\begin{acknowledgements}
The authors would like to thank Cornelius Weig for help with an efficient planar implementation of the \textit{LITMUS} test. Some of the results in this paper have been derived using the \textsc{HEALPix} \citep{gorski-2005} package. We acknowledge the use of the Legacy Archive for Microwave Background Data Analysis (LAMBDA). Support for LAMBDA is provided by the NASA Office of Space Science. Most computations were performed using the \textsc{Sage} software package \citep{sage}. This research was performed in the framework of the DFG Forschergruppe 1254 ``Magnetisation of Interstellar and Intergalactic Media: The Prospects of Low-Frequency Radio Observations''. The idea for this work emerged from the very stimulating discussion with Rodion Stepanov during his visit to Germany, which was supported by the DFG--RFBR grant 08-02-92881. We thank the anonymous referee for constructive comments.
\end{acknowledgements}

\bibliographystyle{myaa}
\bibliography{helicity}

\end{document}

%% file: profilea.tex
\begingroup
  \makeatletter
  \providecommand\color[2][]{%
    \GenericError{(gnuplot) \space\space\space\@spaces}{%
      Package color not loaded in conjunction with
      terminal option `colourtext'%
    }{See the gnuplot documentation for explanation.%
    }{Either use 'blacktext' in gnuplot or load the package
      color.sty in LaTeX.}%
    \renewcommand\color[2][]{}%
  }%
  \providecommand\includegraphics[2][]{%
    \GenericError{(gnuplot) \space\space\space\@spaces}{%
      Package graphicx or graphics not loaded%
    }{See the gnuplot documentation for explanation.%
    }{The gnuplot epslatex terminal needs graphicx.sty or graphics.sty.}%
    \renewcommand\includegraphics[2][]{}%
  }%
  \providecommand\rotatebox[2]{#2}%
  \@ifundefined{ifGPcolor}{%
    \newif\ifGPcolor
    \GPcolorfalse
  }{}%
  \@ifundefined{ifGPblacktext}{%
    \newif\ifGPblacktext
    \GPblacktexttrue
  }{}%
  \let\gplgaddtomacro\g@addto@macro
  \gdef\gplbacktext{}%
  \gdef\gplfronttext{}%
  \makeatother
  \ifGPblacktext
    \def\colorrgb#1{}%
    \def\colorgray#1{}%
  \else
    \ifGPcolor
      \def\colorrgb#1{\color[rgb]{#1}}%
      \def\colorgray#1{\color[gray]{#1}}%
      \expandafter\def\csname LTw\endcsname{\color{white}}%
      \expandafter\def\csname LTb\endcsname{\color{black}}%
      \expandafter\def\csname LTa\endcsname{\color{black}}%
      \expandafter\def\csname LT0\endcsname{\color[rgb]{1,0,0}}%
      \expandafter\def\csname LT1\endcsname{\color[rgb]{0,1,0}}%
      \expandafter\def\csname LT2\endcsname{\color[rgb]{0,0,1}}%
      \expandafter\def\csname LT3\endcsname{\color[rgb]{1,0,1}}%
      \expandafter\def\csname LT4\endcsname{\color[rgb]{0,1,1}}%
      \expandafter\def\csname LT5\endcsname{\color[rgb]{1,1,0}}%
      \expandafter\def\csname LT6\endcsname{\color[rgb]{0,0,0}}%
      \expandafter\def\csname LT7\endcsname{\color[rgb]{1,0.3,0}}%
      \expandafter\def\csname LT8\endcsname{\color[rgb]{0.5,0.5,0.5}}%
    \else
      \def\colorrgb#1{\color{black}}%
      \def\colorgray#1{\color[gray]{#1}}%
      \expandafter\def\csname LTw\endcsname{\color{white}}%
      \expandafter\def\csname LTb\endcsname{\color{black}}%
      \expandafter\def\csname LTa\endcsname{\color{black}}%
      \expandafter\def\csname LT0\endcsname{\color{black}}%
      \expandafter\def\csname LT1\endcsname{\color{black}}%
      \expandafter\def\csname LT2\endcsname{\color{black}}%
      \expandafter\def\csname LT3\endcsname{\color{black}}%
      \expandafter\def\csname LT4\endcsname{\color{black}}%
      \expandafter\def\csname LT5\endcsname{\color{black}}%
      \expandafter\def\csname LT6\endcsname{\color{black}}%
      \expandafter\def\csname LT7\endcsname{\color{black}}%
      \expandafter\def\csname LT8\endcsname{\color{black}}%
    \fi
  \fi
  \setlength{\unitlength}{0.0500bp}%
  \begin{picture}(7200.00,5040.00)%
    \gplgaddtomacro\gplbacktext{%
      \csname LTb\endcsname%
      \put(1254,704){\makebox(0,0)[r]{\strut{} 0}}%
      \put(1254,1213){\makebox(0,0)[r]{\strut{} 20}}%
      \put(1254,1722){\makebox(0,0)[r]{\strut{} 40}}%
      \put(1254,2231){\makebox(0,0)[r]{\strut{} 60}}%
      \put(1254,2740){\makebox(0,0)[r]{\strut{} 80}}%
      \put(1254,3249){\makebox(0,0)[r]{\strut{} 100}}%
      \put(1254,3758){\makebox(0,0)[r]{\strut{} 120}}%
      \put(1254,4267){\makebox(0,0)[r]{\strut{} 140}}%
      \put(1254,4776){\makebox(0,0)[r]{\strut{} 160}}%
      \put(1386,484){\makebox(0,0){\strut{} 0}}%
      \put(2273,484){\makebox(0,0){\strut{} 0.5}}%
      \put(3160,484){\makebox(0,0){\strut{} 1}}%
      \put(4046,484){\makebox(0,0){\strut{} 1.5}}%
      \put(4933,484){\makebox(0,0){\strut{} 2}}%
      \put(5820,484){\makebox(0,0){\strut{} 2.5}}%
      \put(6707,484){\makebox(0,0){\strut{} 3}}%
      \put(484,2740){\rotatebox{90}{\makebox(0,0){\strut{}$p(\vartheta)$}}}%
      \put(4172,154){\makebox(0,0){\strut{}$\vartheta$}}%
    }%
    \gplgaddtomacro\gplfronttext{%
    }%
    \gplbacktext
    \put(0,0){\includegraphics{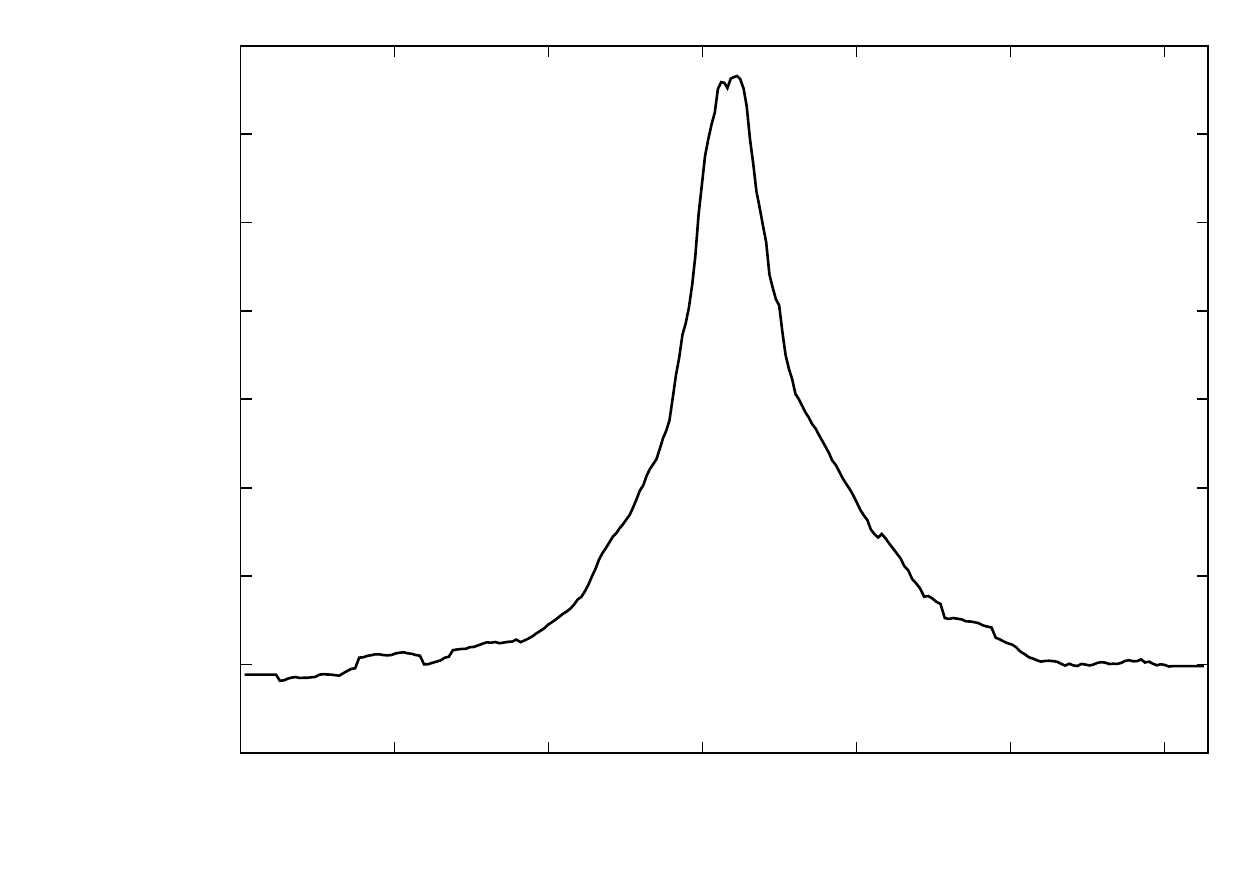}}%
    \gplfronttext
  \end{picture}%
\endgroup

%% file: Cla.tex
\begingroup
  \makeatletter
  \providecommand\color[2][]{%
    \GenericError{(gnuplot) \space\space\space\@spaces}{%
      Package color not loaded in conjunction with
      terminal option `colourtext'%
    }{See the gnuplot documentation for explanation.%
    }{Either use 'blacktext' in gnuplot or load the package
      color.sty in LaTeX.}%
    \renewcommand\color[2][]{}%
  }%
  \providecommand\includegraphics[2][]{%
    \GenericError{(gnuplot) \space\space\space\@spaces}{%
      Package graphicx or graphics not loaded%
    }{See the gnuplot documentation for explanation.%
    }{The gnuplot epslatex terminal needs graphicx.sty or graphics.sty.}%
    \renewcommand\includegraphics[2][]{}%
  }%
  \providecommand\rotatebox[2]{#2}%
  \@ifundefined{ifGPcolor}{%
    \newif\ifGPcolor
    \GPcolorfalse
  }{}%
  \@ifundefined{ifGPblacktext}{%
    \newif\ifGPblacktext
    \GPblacktexttrue
  }{}%
  \let\gplgaddtomacro\g@addto@macro
  \gdef\gplbacktext{}%
  \gdef\gplfronttext{}%
  \makeatother
  \ifGPblacktext
    \def\colorrgb#1{}%
    \def\colorgray#1{}%
  \else
    \ifGPcolor
      \def\colorrgb#1{\color[rgb]{#1}}%
      \def\colorgray#1{\color[gray]{#1}}%
      \expandafter\def\csname LTw\endcsname{\color{white}}%
      \expandafter\def\csname LTb\endcsname{\color{black}}%
      \expandafter\def\csname LTa\endcsname{\color{black}}%
      \expandafter\def\csname LT0\endcsname{\color[rgb]{1,0,0}}%
      \expandafter\def\csname LT1\endcsname{\color[rgb]{0,1,0}}%
      \expandafter\def\csname LT2\endcsname{\color[rgb]{0,0,1}}%
      \expandafter\def\csname LT3\endcsname{\color[rgb]{1,0,1}}%
      \expandafter\def\csname LT4\endcsname{\color[rgb]{0,1,1}}%
      \expandafter\def\csname LT5\endcsname{\color[rgb]{1,1,0}}%
      \expandafter\def\csname LT6\endcsname{\color[rgb]{0,0,0}}%
      \expandafter\def\csname LT7\endcsname{\color[rgb]{1,0.3,0}}%
      \expandafter\def\csname LT8\endcsname{\color[rgb]{0.5,0.5,0.5}}%
    \else
      \def\colorrgb#1{\color{black}}%
      \def\colorgray#1{\color[gray]{#1}}%
      \expandafter\def\csname LTw\endcsname{\color{white}}%
      \expandafter\def\csname LTb\endcsname{\color{black}}%
      \expandafter\def\csname LTa\endcsname{\color{black}}%
      \expandafter\def\csname LT0\endcsname{\color{black}}%
      \expandafter\def\csname LT1\endcsname{\color{black}}%
      \expandafter\def\csname LT2\endcsname{\color{black}}%
      \expandafter\def\csname LT3\endcsname{\color{black}}%
      \expandafter\def\csname LT4\endcsname{\color{black}}%
      \expandafter\def\csname LT5\endcsname{\color{black}}%
      \expandafter\def\csname LT6\endcsname{\color{black}}%
      \expandafter\def\csname LT7\endcsname{\color{black}}%
      \expandafter\def\csname LT8\endcsname{\color{black}}%
    \fi
  \fi
  \setlength{\unitlength}{0.0500bp}%
  \begin{picture}(7200.00,5040.00)%
    \gplgaddtomacro\gplbacktext{%
      \csname LTb\endcsname%
      \put(1650,704){\makebox(0,0)[r]{\strut{} 1e-05}}%
      \put(1650,1518){\makebox(0,0)[r]{\strut{} 0.0001}}%
      \put(1650,2333){\makebox(0,0)[r]{\strut{} 0.001}}%
      \put(1650,3147){\makebox(0,0)[r]{\strut{} 0.01}}%
      \put(1650,3962){\makebox(0,0)[r]{\strut{} 0.1}}%
      \put(1650,4776){\makebox(0,0)[r]{\strut{} 1}}%
      \put(1782,484){\makebox(0,0){\strut{} 1}}%
      \put(4051,484){\makebox(0,0){\strut{} 10}}%
      \put(6320,484){\makebox(0,0){\strut{} 100}}%
      \put(484,2740){\rotatebox{90}{\makebox(0,0){\strut{}$C_l$}}}%
      \put(4370,154){\makebox(0,0){\strut{}$l$}}%
    }%
    \gplgaddtomacro\gplfronttext{%
    }%
    \gplbacktext
    \put(0,0){\includegraphics{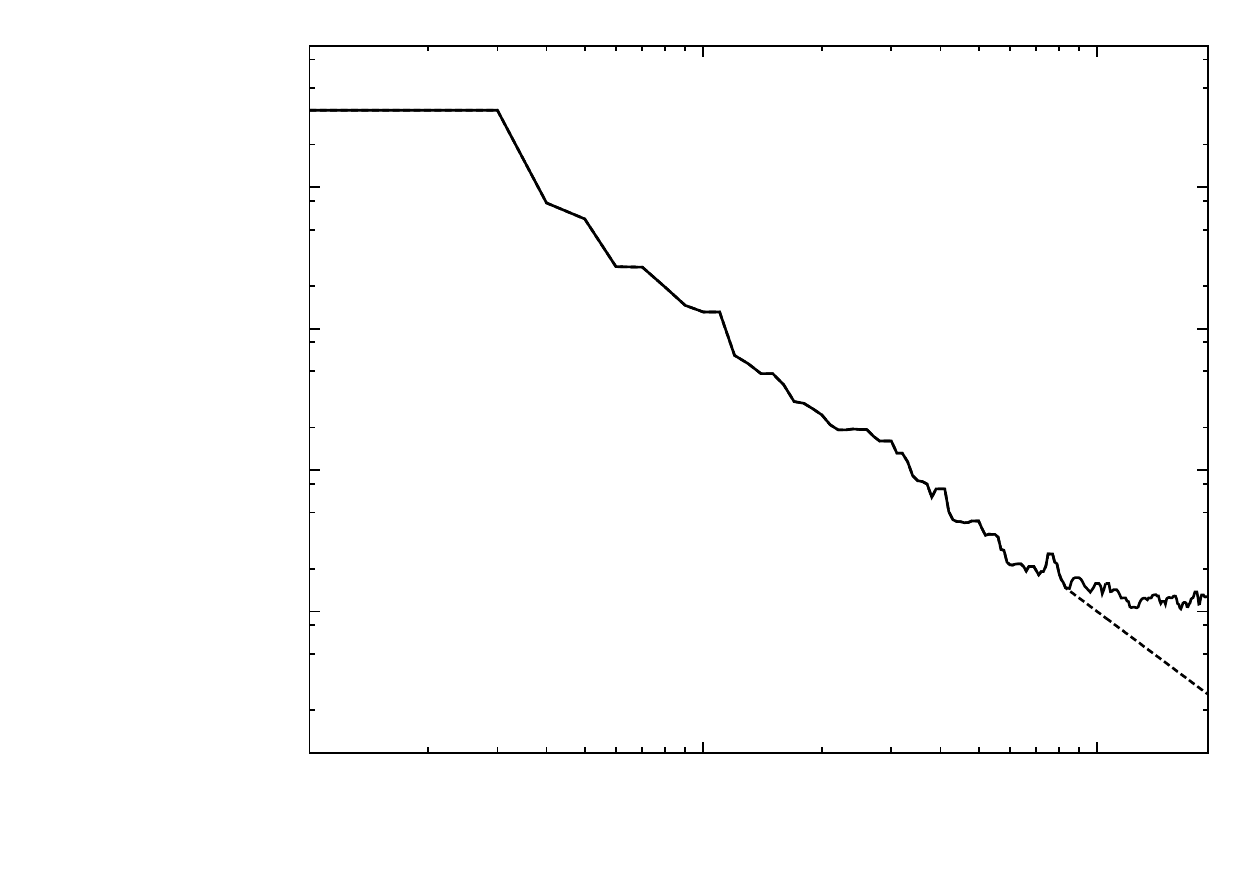}}%
    \gplfronttext
  \end{picture}%
\endgroup

%% file: rotation-obsa.tex
\begingroup
  \makeatletter
  \providecommand\color[2][]{%
    \GenericError{(gnuplot) \space\space\space\@spaces}{%
      Package color not loaded in conjunction with
      terminal option `colourtext'%
    }{See the gnuplot documentation for explanation.%
    }{Either use 'blacktext' in gnuplot or load the package
      color.sty in LaTeX.}%
    \renewcommand\color[2][]{}%
  }%
  \providecommand\includegraphics[2][]{%
    \GenericError{(gnuplot) \space\space\space\@spaces}{%
      Package graphicx or graphics not loaded%
    }{See the gnuplot documentation for explanation.%
    }{The gnuplot epslatex terminal needs graphicx.sty or graphics.sty.}%
    \renewcommand\includegraphics[2][]{}%
  }%
  \providecommand\rotatebox[2]{#2}%
  \@ifundefined{ifGPcolor}{%
    \newif\ifGPcolor
    \GPcolorfalse
  }{}%
  \@ifundefined{ifGPblacktext}{%
    \newif\ifGPblacktext
    \GPblacktexttrue
  }{}%
  \let\gplgaddtomacro\g@addto@macro
  \gdef\gplbacktext{}%
  \gdef\gplfronttext{}%
  \makeatother
  \ifGPblacktext
    \def\colorrgb#1{}%
    \def\colorgray#1{}%
  \else
    \ifGPcolor
      \def\colorrgb#1{\color[rgb]{#1}}%
      \def\colorgray#1{\color[gray]{#1}}%
      \expandafter\def\csname LTw\endcsname{\color{white}}%
      \expandafter\def\csname LTb\endcsname{\color{black}}%
      \expandafter\def\csname LTa\endcsname{\color{black}}%
      \expandafter\def\csname LT0\endcsname{\color[rgb]{1,0,0}}%
      \expandafter\def\csname LT1\endcsname{\color[rgb]{0,1,0}}%
      \expandafter\def\csname LT2\endcsname{\color[rgb]{0,0,1}}%
      \expandafter\def\csname LT3\endcsname{\color[rgb]{1,0,1}}%
      \expandafter\def\csname LT4\endcsname{\color[rgb]{0,1,1}}%
      \expandafter\def\csname LT5\endcsname{\color[rgb]{1,1,0}}%
      \expandafter\def\csname LT6\endcsname{\color[rgb]{0,0,0}}%
      \expandafter\def\csname LT7\endcsname{\color[rgb]{1,0.3,0}}%
      \expandafter\def\csname LT8\endcsname{\color[rgb]{0.5,0.5,0.5}}%
    \else
      \def\colorrgb#1{\color{black}}%
      \def\colorgray#1{\color[gray]{#1}}%
      \expandafter\def\csname LTw\endcsname{\color{white}}%
      \expandafter\def\csname LTb\endcsname{\color{black}}%
      \expandafter\def\csname LTa\endcsname{\color{black}}%
      \expandafter\def\csname LT0\endcsname{\color{black}}%
      \expandafter\def\csname LT1\endcsname{\color{black}}%
      \expandafter\def\csname LT2\endcsname{\color{black}}%
      \expandafter\def\csname LT3\endcsname{\color{black}}%
      \expandafter\def\csname LT4\endcsname{\color{black}}%
      \expandafter\def\csname LT5\endcsname{\color{black}}%
      \expandafter\def\csname LT6\endcsname{\color{black}}%
      \expandafter\def\csname LT7\endcsname{\color{black}}%
      \expandafter\def\csname LT8\endcsname{\color{black}}%
    \fi
  \fi
  \setlength{\unitlength}{0.0500bp}%
  \begin{picture}(7200.00,5040.00)%
    \gplgaddtomacro\gplbacktext{%
      \csname LTb\endcsname%
      \put(1254,975){\makebox(0,0)[r]{\strut{}-1}}%
      \put(1254,1518){\makebox(0,0)[r]{\strut{}-0.8}}%
      \put(1254,2061){\makebox(0,0)[r]{\strut{}-0.6}}%
      \put(1254,2604){\makebox(0,0)[r]{\strut{}-0.4}}%
      \put(1254,3147){\makebox(0,0)[r]{\strut{}-0.2}}%
      \put(1254,3690){\makebox(0,0)[r]{\strut{} 0}}%
      \put(1254,4233){\makebox(0,0)[r]{\strut{} 0.2}}%
      \put(1254,4776){\makebox(0,0)[r]{\strut{} 0.4}}%
      \put(1512,484){\makebox(0,0){\strut{}-3}}%
      \put(2398,484){\makebox(0,0){\strut{}-2}}%
      \put(3285,484){\makebox(0,0){\strut{}-1}}%
      \put(4172,484){\makebox(0,0){\strut{} 0}}%
      \put(5059,484){\makebox(0,0){\strut{} 1}}%
      \put(5946,484){\makebox(0,0){\strut{} 2}}%
      \put(6832,484){\makebox(0,0){\strut{} 3}}%
      \put(484,2740){\rotatebox{90}{\makebox(0,0){\strut{}$\mathrm{Re}\left<GP^*\right>_{\mathcal{S}^2}$}}}%
      \put(4172,154){\makebox(0,0){\strut{}$\beta$}}%
    }%
    \gplgaddtomacro\gplfronttext{%
    }%
    \gplbacktext
    \put(0,0){\includegraphics{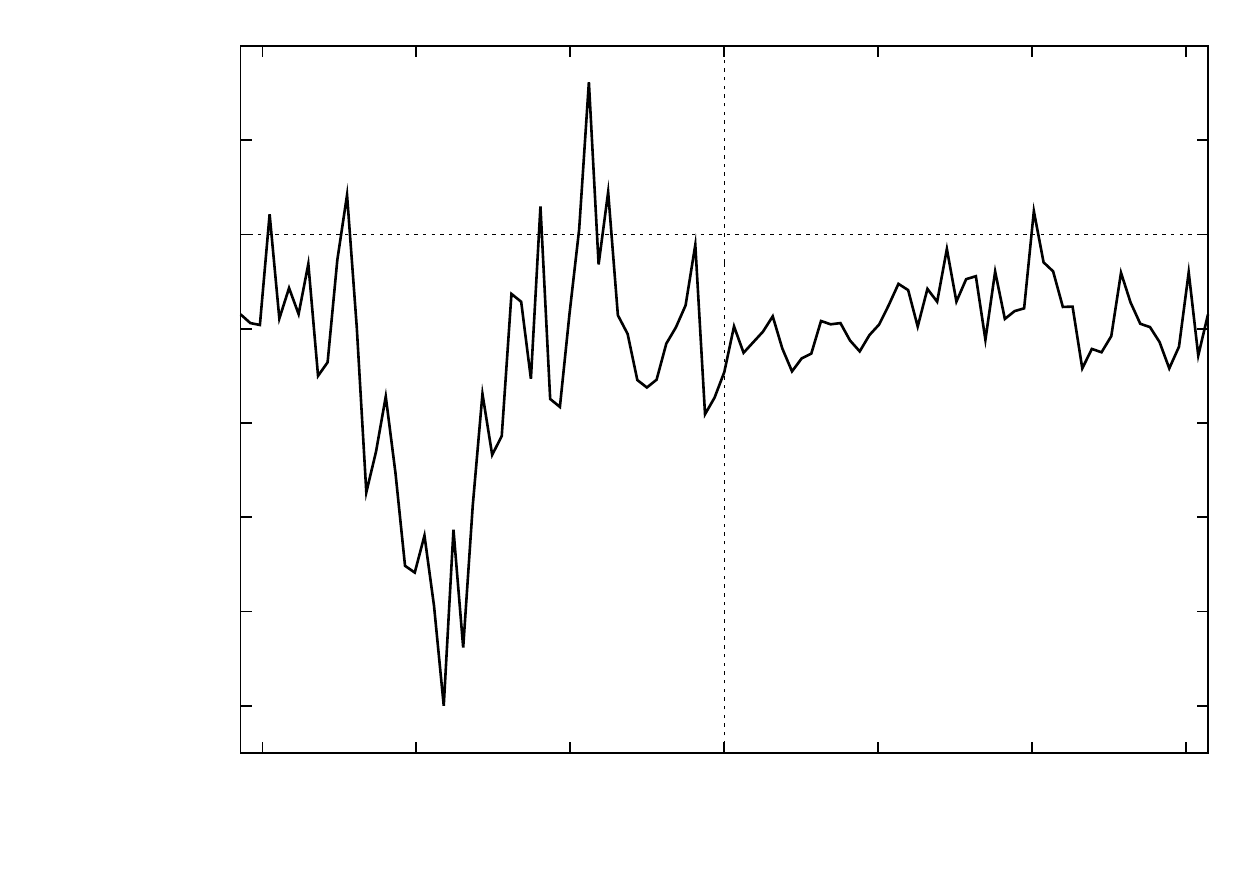}}%
    \gplfronttext
  \end{picture}%
\endgroup

%% file: rotation-hel-nohel-noela.tex
\begingroup
  \makeatletter
  \providecommand\color[2][]{%
    \GenericError{(gnuplot) \space\space\space\@spaces}{%
      Package color not loaded in conjunction with
      terminal option `colourtext'%
    }{See the gnuplot documentation for explanation.%
    }{Either use 'blacktext' in gnuplot or load the package
      color.sty in LaTeX.}%
    \renewcommand\color[2][]{}%
  }%
  \providecommand\includegraphics[2][]{%
    \GenericError{(gnuplot) \space\space\space\@spaces}{%
      Package graphicx or graphics not loaded%
    }{See the gnuplot documentation for explanation.%
    }{The gnuplot epslatex terminal needs graphicx.sty or graphics.sty.}%
    \renewcommand\includegraphics[2][]{}%
  }%
  \providecommand\rotatebox[2]{#2}%
  \@ifundefined{ifGPcolor}{%
    \newif\ifGPcolor
    \GPcolorfalse
  }{}%
  \@ifundefined{ifGPblacktext}{%
    \newif\ifGPblacktext
    \GPblacktexttrue
  }{}%
  \let\gplgaddtomacro\g@addto@macro
  \gdef\gplbacktext{}%
  \gdef\gplfronttext{}%
  \makeatother
  \ifGPblacktext
    \def\colorrgb#1{}%
    \def\colorgray#1{}%
  \else
    \ifGPcolor
      \def\colorrgb#1{\color[rgb]{#1}}%
      \def\colorgray#1{\color[gray]{#1}}%
      \expandafter\def\csname LTw\endcsname{\color{white}}%
      \expandafter\def\csname LTb\endcsname{\color{black}}%
      \expandafter\def\csname LTa\endcsname{\color{black}}%
      \expandafter\def\csname LT0\endcsname{\color[rgb]{1,0,0}}%
      \expandafter\def\csname LT1\endcsname{\color[rgb]{0,1,0}}%
      \expandafter\def\csname LT2\endcsname{\color[rgb]{0,0,1}}%
      \expandafter\def\csname LT3\endcsname{\color[rgb]{1,0,1}}%
      \expandafter\def\csname LT4\endcsname{\color[rgb]{0,1,1}}%
      \expandafter\def\csname LT5\endcsname{\color[rgb]{1,1,0}}%
      \expandafter\def\csname LT6\endcsname{\color[rgb]{0,0,0}}%
      \expandafter\def\csname LT7\endcsname{\color[rgb]{1,0.3,0}}%
      \expandafter\def\csname LT8\endcsname{\color[rgb]{0.5,0.5,0.5}}%
    \else
      \def\colorrgb#1{\color{black}}%
      \def\colorgray#1{\color[gray]{#1}}%
      \expandafter\def\csname LTw\endcsname{\color{white}}%
      \expandafter\def\csname LTb\endcsname{\color{black}}%
      \expandafter\def\csname LTa\endcsname{\color{black}}%
      \expandafter\def\csname LT0\endcsname{\color{black}}%
      \expandafter\def\csname LT1\endcsname{\color{black}}%
      \expandafter\def\csname LT2\endcsname{\color{black}}%
      \expandafter\def\csname LT3\endcsname{\color{black}}%
      \expandafter\def\csname LT4\endcsname{\color{black}}%
      \expandafter\def\csname LT5\endcsname{\color{black}}%
      \expandafter\def\csname LT6\endcsname{\color{black}}%
      \expandafter\def\csname LT7\endcsname{\color{black}}%
      \expandafter\def\csname LT8\endcsname{\color{black}}%
    \fi
  \fi
  \setlength{\unitlength}{0.0500bp}%
  \begin{picture}(7200.00,5040.00)%
    \gplgaddtomacro\gplbacktext{%
      \csname LTb\endcsname%
      \put(1254,704){\makebox(0,0)[r]{\strut{}-0.4}}%
      \put(1254,1247){\makebox(0,0)[r]{\strut{}-0.2}}%
      \put(1254,1790){\makebox(0,0)[r]{\strut{} 0}}%
      \put(1254,2333){\makebox(0,0)[r]{\strut{} 0.2}}%
      \put(1254,2876){\makebox(0,0)[r]{\strut{} 0.4}}%
      \put(1254,3419){\makebox(0,0)[r]{\strut{} 0.6}}%
      \put(1254,3962){\makebox(0,0)[r]{\strut{} 0.8}}%
      \put(1254,4505){\makebox(0,0)[r]{\strut{} 1}}%
      \put(1512,484){\makebox(0,0){\strut{}-3}}%
      \put(2398,484){\makebox(0,0){\strut{}-2}}%
      \put(3285,484){\makebox(0,0){\strut{}-1}}%
      \put(4172,484){\makebox(0,0){\strut{} 0}}%
      \put(5059,484){\makebox(0,0){\strut{} 1}}%
      \put(5946,484){\makebox(0,0){\strut{} 2}}%
      \put(6832,484){\makebox(0,0){\strut{} 3}}%
      \put(484,2740){\rotatebox{90}{\makebox(0,0){\strut{}$\mathrm{Re}\left<GP^*\right>_{\mathcal{S}^2}$}}}%
      \put(4172,154){\makebox(0,0){\strut{}$\beta$}}%
    }%
    \gplgaddtomacro\gplfronttext{%
    }%
    \gplbacktext
    \put(0,0){\includegraphics{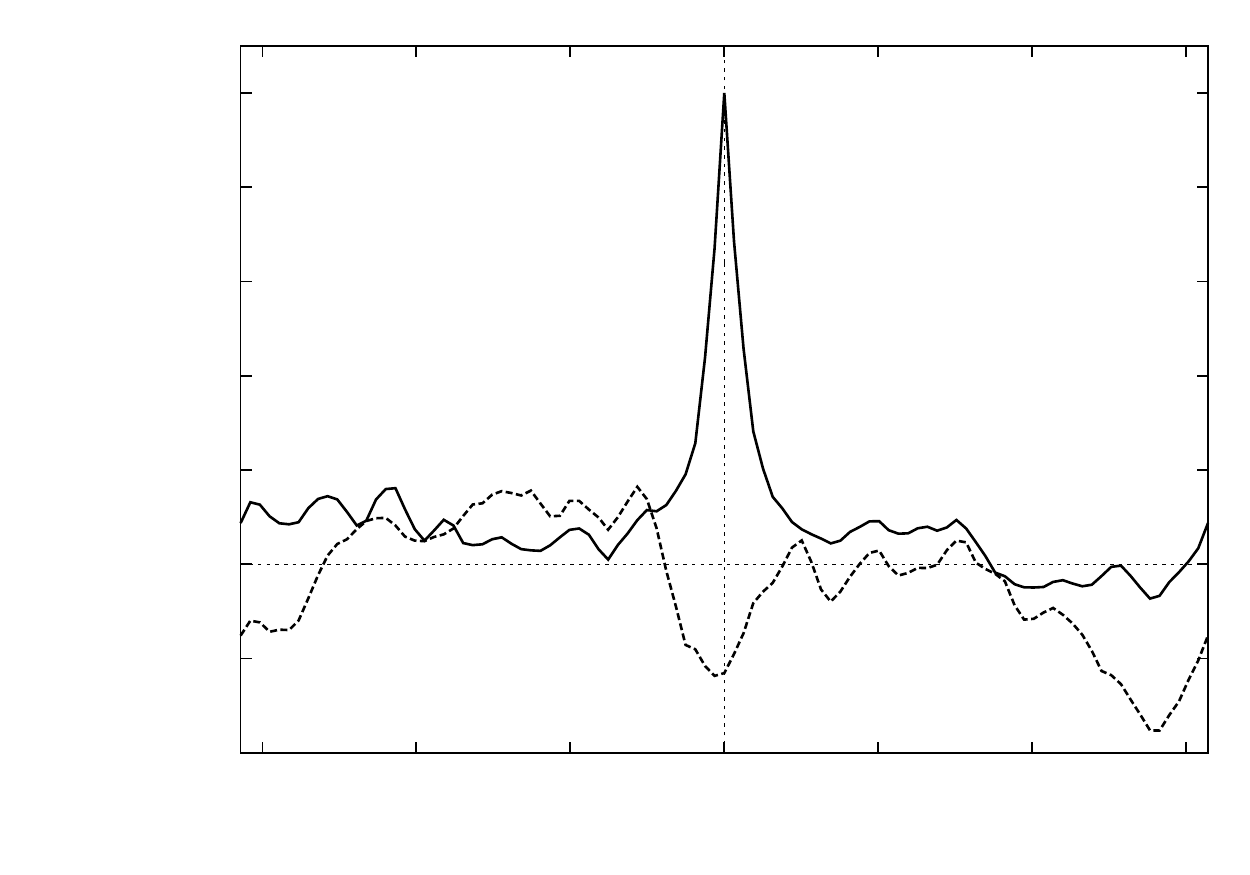}}%
    \gplfronttext
  \end{picture}%
\endgroup

%% file: rotation-hel-nohel-noel-small-scalea.tex
\begingroup
  \makeatletter
  \providecommand\color[2][]{%
    \GenericError{(gnuplot) \space\space\space\@spaces}{%
      Package color not loaded in conjunction with
      terminal option `colourtext'%
    }{See the gnuplot documentation for explanation.%
    }{Either use 'blacktext' in gnuplot or load the package
      color.sty in LaTeX.}%
    \renewcommand\color[2][]{}%
  }%
  \providecommand\includegraphics[2][]{%
    \GenericError{(gnuplot) \space\space\space\@spaces}{%
      Package graphicx or graphics not loaded%
    }{See the gnuplot documentation for explanation.%
    }{The gnuplot epslatex terminal needs graphicx.sty or graphics.sty.}%
    \renewcommand\includegraphics[2][]{}%
  }%
  \providecommand\rotatebox[2]{#2}%
  \@ifundefined{ifGPcolor}{%
    \newif\ifGPcolor
    \GPcolorfalse
  }{}%
  \@ifundefined{ifGPblacktext}{%
    \newif\ifGPblacktext
    \GPblacktexttrue
  }{}%
  \let\gplgaddtomacro\g@addto@macro
  \gdef\gplbacktext{}%
  \gdef\gplfronttext{}%
  \makeatother
  \ifGPblacktext
    \def\colorrgb#1{}%
    \def\colorgray#1{}%
  \else
    \ifGPcolor
      \def\colorrgb#1{\color[rgb]{#1}}%
      \def\colorgray#1{\color[gray]{#1}}%
      \expandafter\def\csname LTw\endcsname{\color{white}}%
      \expandafter\def\csname LTb\endcsname{\color{black}}%
      \expandafter\def\csname LTa\endcsname{\color{black}}%
      \expandafter\def\csname LT0\endcsname{\color[rgb]{1,0,0}}%
      \expandafter\def\csname LT1\endcsname{\color[rgb]{0,1,0}}%
      \expandafter\def\csname LT2\endcsname{\color[rgb]{0,0,1}}%
      \expandafter\def\csname LT3\endcsname{\color[rgb]{1,0,1}}%
      \expandafter\def\csname LT4\endcsname{\color[rgb]{0,1,1}}%
      \expandafter\def\csname LT5\endcsname{\color[rgb]{1,1,0}}%
      \expandafter\def\csname LT6\endcsname{\color[rgb]{0,0,0}}%
      \expandafter\def\csname LT7\endcsname{\color[rgb]{1,0.3,0}}%
      \expandafter\def\csname LT8\endcsname{\color[rgb]{0.5,0.5,0.5}}%
    \else
      \def\colorrgb#1{\color{black}}%
      \def\colorgray#1{\color[gray]{#1}}%
      \expandafter\def\csname LTw\endcsname{\color{white}}%
      \expandafter\def\csname LTb\endcsname{\color{black}}%
      \expandafter\def\csname LTa\endcsname{\color{black}}%
      \expandafter\def\csname LT0\endcsname{\color{black}}%
      \expandafter\def\csname LT1\endcsname{\color{black}}%
      \expandafter\def\csname LT2\endcsname{\color{black}}%
      \expandafter\def\csname LT3\endcsname{\color{black}}%
      \expandafter\def\csname LT4\endcsname{\color{black}}%
      \expandafter\def\csname LT5\endcsname{\color{black}}%
      \expandafter\def\csname LT6\endcsname{\color{black}}%
      \expandafter\def\csname LT7\endcsname{\color{black}}%
      \expandafter\def\csname LT8\endcsname{\color{black}}%
    \fi
  \fi
  \setlength{\unitlength}{0.0500bp}%
  \begin{picture}(7200.00,5040.00)%
    \gplgaddtomacro\gplbacktext{%
      \csname LTb\endcsname%
      \put(1254,704){\makebox(0,0)[r]{\strut{}-0.4}}%
      \put(1254,1247){\makebox(0,0)[r]{\strut{}-0.2}}%
      \put(1254,1790){\makebox(0,0)[r]{\strut{} 0}}%
      \put(1254,2333){\makebox(0,0)[r]{\strut{} 0.2}}%
      \put(1254,2876){\makebox(0,0)[r]{\strut{} 0.4}}%
      \put(1254,3419){\makebox(0,0)[r]{\strut{} 0.6}}%
      \put(1254,3962){\makebox(0,0)[r]{\strut{} 0.8}}%
      \put(1254,4505){\makebox(0,0)[r]{\strut{} 1}}%
      \put(1512,484){\makebox(0,0){\strut{}-3}}%
      \put(2398,484){\makebox(0,0){\strut{}-2}}%
      \put(3285,484){\makebox(0,0){\strut{}-1}}%
      \put(4172,484){\makebox(0,0){\strut{} 0}}%
      \put(5059,484){\makebox(0,0){\strut{} 1}}%
      \put(5946,484){\makebox(0,0){\strut{} 2}}%
      \put(6832,484){\makebox(0,0){\strut{} 3}}%
      \put(484,2740){\rotatebox{90}{\makebox(0,0){\strut{}$\textrm{Re}\left(\sum\limits_{l=25}^{l_\mathrm{max}}\sum\limits_{m=-l}^{l}G_{lm}P_{lm}^*\right)$}}}%
      \put(4172,154){\makebox(0,0){\strut{}$\beta$}}%
    }%
    \gplgaddtomacro\gplfronttext{%
    }%
    \gplbacktext
    \put(0,0){\includegraphics{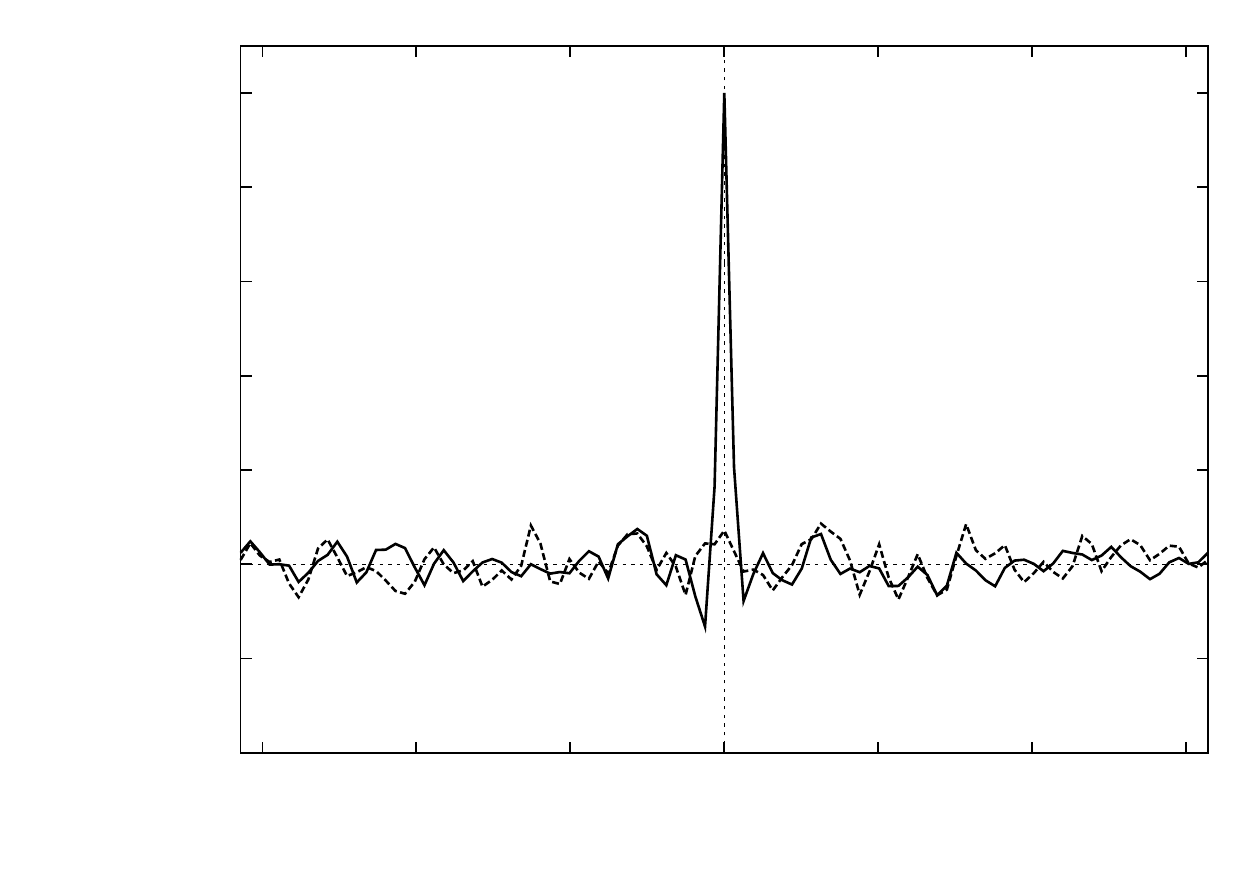}}%
    \gplfronttext
  \end{picture}%
\endgroup

%% file: rotation-obs-small-scalea.tex
\begingroup
  \makeatletter
  \providecommand\color[2][]{%
    \GenericError{(gnuplot) \space\space\space\@spaces}{%
      Package color not loaded in conjunction with
      terminal option `colourtext'%
    }{See the gnuplot documentation for explanation.%
    }{Either use 'blacktext' in gnuplot or load the package
      color.sty in LaTeX.}%
    \renewcommand\color[2][]{}%
  }%
  \providecommand\includegraphics[2][]{%
    \GenericError{(gnuplot) \space\space\space\@spaces}{%
      Package graphicx or graphics not loaded%
    }{See the gnuplot documentation for explanation.%
    }{The gnuplot epslatex terminal needs graphicx.sty or graphics.sty.}%
    \renewcommand\includegraphics[2][]{}%
  }%
  \providecommand\rotatebox[2]{#2}%
  \@ifundefined{ifGPcolor}{%
    \newif\ifGPcolor
    \GPcolorfalse
  }{}%
  \@ifundefined{ifGPblacktext}{%
    \newif\ifGPblacktext
    \GPblacktexttrue
  }{}%
  \let\gplgaddtomacro\g@addto@macro
  \gdef\gplbacktext{}%
  \gdef\gplfronttext{}%
  \makeatother
  \ifGPblacktext
    \def\colorrgb#1{}%
    \def\colorgray#1{}%
  \else
    \ifGPcolor
      \def\colorrgb#1{\color[rgb]{#1}}%
      \def\colorgray#1{\color[gray]{#1}}%
      \expandafter\def\csname LTw\endcsname{\color{white}}%
      \expandafter\def\csname LTb\endcsname{\color{black}}%
      \expandafter\def\csname LTa\endcsname{\color{black}}%
      \expandafter\def\csname LT0\endcsname{\color[rgb]{1,0,0}}%
      \expandafter\def\csname LT1\endcsname{\color[rgb]{0,1,0}}%
      \expandafter\def\csname LT2\endcsname{\color[rgb]{0,0,1}}%
      \expandafter\def\csname LT3\endcsname{\color[rgb]{1,0,1}}%
      \expandafter\def\csname LT4\endcsname{\color[rgb]{0,1,1}}%
      \expandafter\def\csname LT5\endcsname{\color[rgb]{1,1,0}}%
      \expandafter\def\csname LT6\endcsname{\color[rgb]{0,0,0}}%
      \expandafter\def\csname LT7\endcsname{\color[rgb]{1,0.3,0}}%
      \expandafter\def\csname LT8\endcsname{\color[rgb]{0.5,0.5,0.5}}%
    \else
      \def\colorrgb#1{\color{black}}%
      \def\colorgray#1{\color[gray]{#1}}%
      \expandafter\def\csname LTw\endcsname{\color{white}}%
      \expandafter\def\csname LTb\endcsname{\color{black}}%
      \expandafter\def\csname LTa\endcsname{\color{black}}%
      \expandafter\def\csname LT0\endcsname{\color{black}}%
      \expandafter\def\csname LT1\endcsname{\color{black}}%
      \expandafter\def\csname LT2\endcsname{\color{black}}%
      \expandafter\def\csname LT3\endcsname{\color{black}}%
      \expandafter\def\csname LT4\endcsname{\color{black}}%
      \expandafter\def\csname LT5\endcsname{\color{black}}%
      \expandafter\def\csname LT6\endcsname{\color{black}}%
      \expandafter\def\csname LT7\endcsname{\color{black}}%
      \expandafter\def\csname LT8\endcsname{\color{black}}%
    \fi
  \fi
  \setlength{\unitlength}{0.0500bp}%
  \begin{picture}(7200.00,5040.00)%
    \gplgaddtomacro\gplbacktext{%
      \csname LTb\endcsname%
      \put(1254,704){\makebox(0,0)[r]{\strut{}-1}}%
      \put(1254,1674){\makebox(0,0)[r]{\strut{}-0.5}}%
      \put(1254,2643){\makebox(0,0)[r]{\strut{} 0}}%
      \put(1254,3613){\makebox(0,0)[r]{\strut{} 0.5}}%
      \put(1254,4582){\makebox(0,0)[r]{\strut{} 1}}%
      \put(1512,484){\makebox(0,0){\strut{}-3}}%
      \put(2398,484){\makebox(0,0){\strut{}-2}}%
      \put(3285,484){\makebox(0,0){\strut{}-1}}%
      \put(4172,484){\makebox(0,0){\strut{} 0}}%
      \put(5059,484){\makebox(0,0){\strut{} 1}}%
      \put(5946,484){\makebox(0,0){\strut{} 2}}%
      \put(6832,484){\makebox(0,0){\strut{} 3}}%
      \put(484,2740){\rotatebox{90}{\makebox(0,0){\strut{}$\textrm{Re}\left(\sum\limits_{l=25}^{l_\mathrm{max}}\sum\limits_{m=-l}^{l}G_{lm}P_{lm}^*\right)$}}}%
      \put(4172,154){\makebox(0,0){\strut{}$\beta$}}%
    }%
    \gplgaddtomacro\gplfronttext{%
    }%
    \gplbacktext
    \put(0,0){\includegraphics{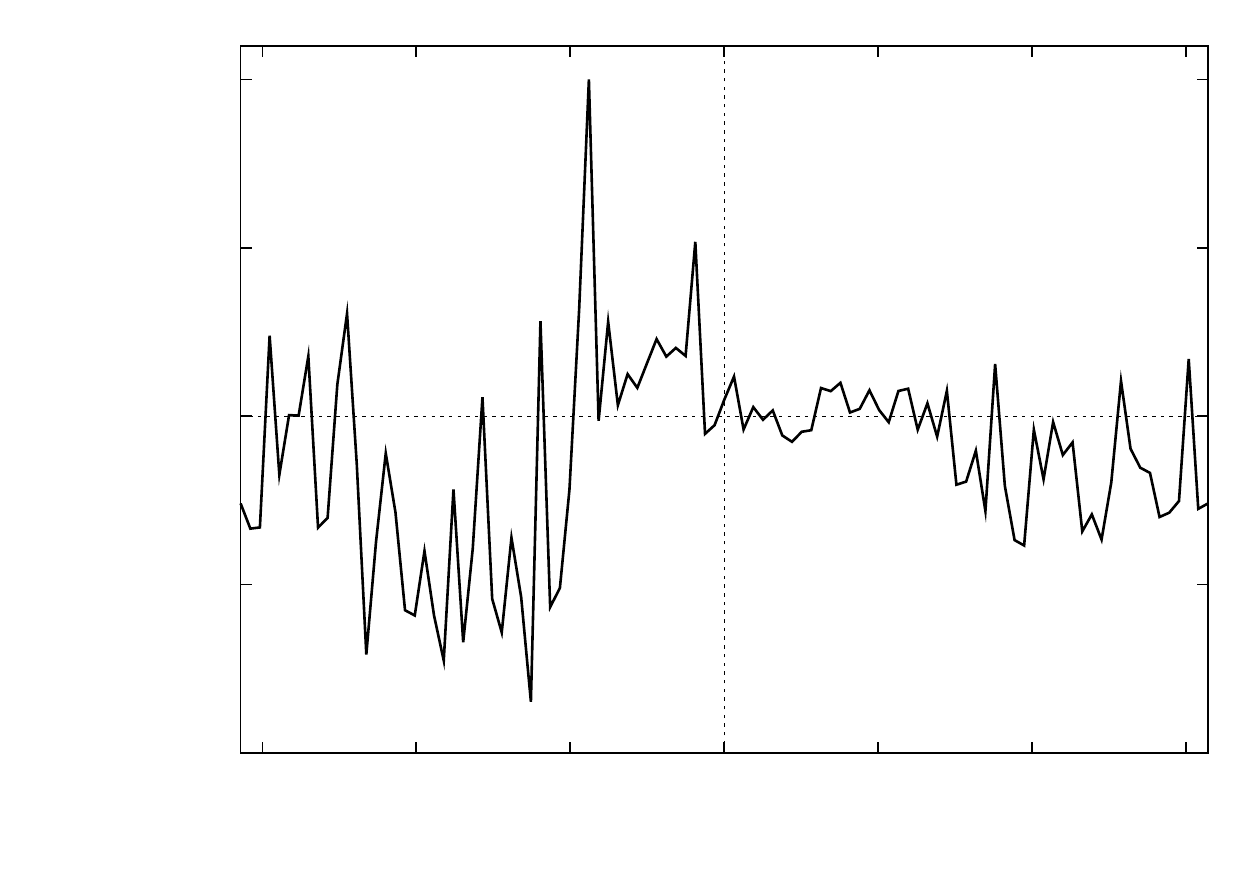}}%
    \gplfronttext
  \end{picture}%
\endgroup

%% file: rotation-WMAP-noela.tex
\begingroup
  \makeatletter
  \providecommand\color[2][]{%
    \GenericError{(gnuplot) \space\space\space\@spaces}{%
      Package color not loaded in conjunction with
      terminal option `colourtext'%
    }{See the gnuplot documentation for explanation.%
    }{Either use 'blacktext' in gnuplot or load the package
      color.sty in LaTeX.}%
    \renewcommand\color[2][]{}%
  }%
  \providecommand\includegraphics[2][]{%
    \GenericError{(gnuplot) \space\space\space\@spaces}{%
      Package graphicx or graphics not loaded%
    }{See the gnuplot documentation for explanation.%
    }{The gnuplot epslatex terminal needs graphicx.sty or graphics.sty.}%
    \renewcommand\includegraphics[2][]{}%
  }%
  \providecommand\rotatebox[2]{#2}%
  \@ifundefined{ifGPcolor}{%
    \newif\ifGPcolor
    \GPcolorfalse
  }{}%
  \@ifundefined{ifGPblacktext}{%
    \newif\ifGPblacktext
    \GPblacktexttrue
  }{}%
  \let\gplgaddtomacro\g@addto@macro
  \gdef\gplbacktext{}%
  \gdef\gplfronttext{}%
  \makeatother
  \ifGPblacktext
    \def\colorrgb#1{}%
    \def\colorgray#1{}%
  \else
    \ifGPcolor
      \def\colorrgb#1{\color[rgb]{#1}}%
      \def\colorgray#1{\color[gray]{#1}}%
      \expandafter\def\csname LTw\endcsname{\color{white}}%
      \expandafter\def\csname LTb\endcsname{\color{black}}%
      \expandafter\def\csname LTa\endcsname{\color{black}}%
      \expandafter\def\csname LT0\endcsname{\color[rgb]{1,0,0}}%
      \expandafter\def\csname LT1\endcsname{\color[rgb]{0,1,0}}%
      \expandafter\def\csname LT2\endcsname{\color[rgb]{0,0,1}}%
      \expandafter\def\csname LT3\endcsname{\color[rgb]{1,0,1}}%
      \expandafter\def\csname LT4\endcsname{\color[rgb]{0,1,1}}%
      \expandafter\def\csname LT5\endcsname{\color[rgb]{1,1,0}}%
      \expandafter\def\csname LT6\endcsname{\color[rgb]{0,0,0}}%
      \expandafter\def\csname LT7\endcsname{\color[rgb]{1,0.3,0}}%
      \expandafter\def\csname LT8\endcsname{\color[rgb]{0.5,0.5,0.5}}%
    \else
      \def\colorrgb#1{\color{black}}%
      \def\colorgray#1{\color[gray]{#1}}%
      \expandafter\def\csname LTw\endcsname{\color{white}}%
      \expandafter\def\csname LTb\endcsname{\color{black}}%
      \expandafter\def\csname LTa\endcsname{\color{black}}%
      \expandafter\def\csname LT0\endcsname{\color{black}}%
      \expandafter\def\csname LT1\endcsname{\color{black}}%
      \expandafter\def\csname LT2\endcsname{\color{black}}%
      \expandafter\def\csname LT3\endcsname{\color{black}}%
      \expandafter\def\csname LT4\endcsname{\color{black}}%
      \expandafter\def\csname LT5\endcsname{\color{black}}%
      \expandafter\def\csname LT6\endcsname{\color{black}}%
      \expandafter\def\csname LT7\endcsname{\color{black}}%
      \expandafter\def\csname LT8\endcsname{\color{black}}%
    \fi
  \fi
  \setlength{\unitlength}{0.0500bp}%
  \begin{picture}(7200.00,5040.00)%
    \gplgaddtomacro\gplbacktext{%
      \csname LTb\endcsname%
      \put(1254,1111){\makebox(0,0)[r]{\strut{} 0.2}}%
      \put(1254,1926){\makebox(0,0)[r]{\strut{} 0.4}}%
      \put(1254,2740){\makebox(0,0)[r]{\strut{} 0.6}}%
      \put(1254,3554){\makebox(0,0)[r]{\strut{} 0.8}}%
      \put(1254,4369){\makebox(0,0)[r]{\strut{} 1}}%
      \put(1512,484){\makebox(0,0){\strut{}-3}}%
      \put(2398,484){\makebox(0,0){\strut{}-2}}%
      \put(3285,484){\makebox(0,0){\strut{}-1}}%
      \put(4172,484){\makebox(0,0){\strut{} 0}}%
      \put(5059,484){\makebox(0,0){\strut{} 1}}%
      \put(5946,484){\makebox(0,0){\strut{} 2}}%
      \put(6832,484){\makebox(0,0){\strut{} 3}}%
      \put(484,2740){\rotatebox{90}{\makebox(0,0){\strut{}$\mathrm{Re}\left<GP^*\right>_{\mathcal{S}^2}$}}}%
      \put(4172,154){\makebox(0,0){\strut{}$\beta$}}%
    }%
    \gplgaddtomacro\gplfronttext{%
    }%
    \gplbacktext
    \put(0,0){\includegraphics{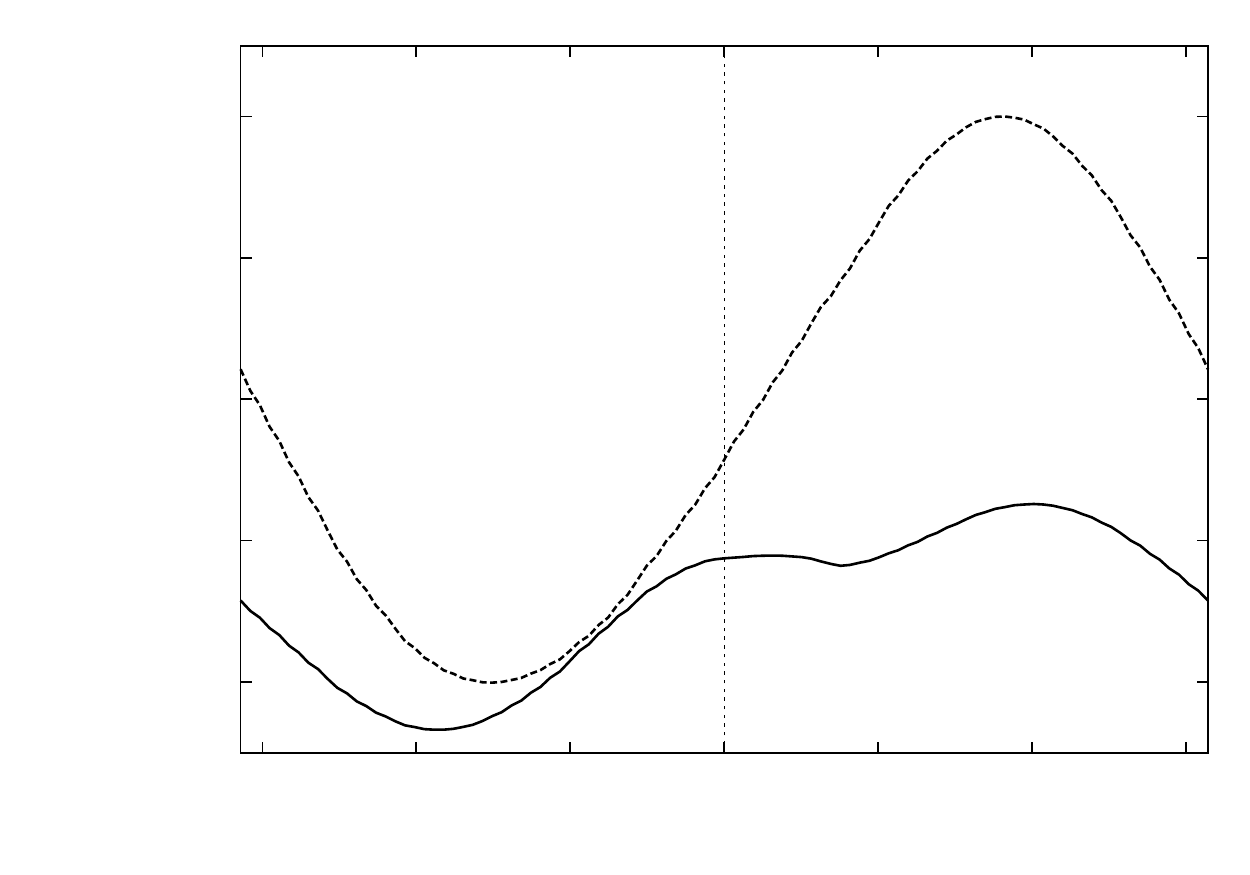}}%
    \gplfronttext
  \end{picture}%
\endgroup

%% file: rotation-hel-nohela.tex
\begingroup
  \makeatletter
  \providecommand\color[2][]{%
    \GenericError{(gnuplot) \space\space\space\@spaces}{%
      Package color not loaded in conjunction with
      terminal option `colourtext'%
    }{See the gnuplot documentation for explanation.%
    }{Either use 'blacktext' in gnuplot or load the package
      color.sty in LaTeX.}%
    \renewcommand\color[2][]{}%
  }%
  \providecommand\includegraphics[2][]{%
    \GenericError{(gnuplot) \space\space\space\@spaces}{%
      Package graphicx or graphics not loaded%
    }{See the gnuplot documentation for explanation.%
    }{The gnuplot epslatex terminal needs graphicx.sty or graphics.sty.}%
    \renewcommand\includegraphics[2][]{}%
  }%
  \providecommand\rotatebox[2]{#2}%
  \@ifundefined{ifGPcolor}{%
    \newif\ifGPcolor
    \GPcolorfalse
  }{}%
  \@ifundefined{ifGPblacktext}{%
    \newif\ifGPblacktext
    \GPblacktexttrue
  }{}%
  \let\gplgaddtomacro\g@addto@macro
  \gdef\gplbacktext{}%
  \gdef\gplfronttext{}%
  \makeatother
  \ifGPblacktext
    \def\colorrgb#1{}%
    \def\colorgray#1{}%
  \else
    \ifGPcolor
      \def\colorrgb#1{\color[rgb]{#1}}%
      \def\colorgray#1{\color[gray]{#1}}%
      \expandafter\def\csname LTw\endcsname{\color{white}}%
      \expandafter\def\csname LTb\endcsname{\color{black}}%
      \expandafter\def\csname LTa\endcsname{\color{black}}%
      \expandafter\def\csname LT0\endcsname{\color[rgb]{1,0,0}}%
      \expandafter\def\csname LT1\endcsname{\color[rgb]{0,1,0}}%
      \expandafter\def\csname LT2\endcsname{\color[rgb]{0,0,1}}%
      \expandafter\def\csname LT3\endcsname{\color[rgb]{1,0,1}}%
      \expandafter\def\csname LT4\endcsname{\color[rgb]{0,1,1}}%
      \expandafter\def\csname LT5\endcsname{\color[rgb]{1,1,0}}%
      \expandafter\def\csname LT6\endcsname{\color[rgb]{0,0,0}}%
      \expandafter\def\csname LT7\endcsname{\color[rgb]{1,0.3,0}}%
      \expandafter\def\csname LT8\endcsname{\color[rgb]{0.5,0.5,0.5}}%
    \else
      \def\colorrgb#1{\color{black}}%
      \def\colorgray#1{\color[gray]{#1}}%
      \expandafter\def\csname LTw\endcsname{\color{white}}%
      \expandafter\def\csname LTb\endcsname{\color{black}}%
      \expandafter\def\csname LTa\endcsname{\color{black}}%
      \expandafter\def\csname LT0\endcsname{\color{black}}%
      \expandafter\def\csname LT1\endcsname{\color{black}}%
      \expandafter\def\csname LT2\endcsname{\color{black}}%
      \expandafter\def\csname LT3\endcsname{\color{black}}%
      \expandafter\def\csname LT4\endcsname{\color{black}}%
      \expandafter\def\csname LT5\endcsname{\color{black}}%
      \expandafter\def\csname LT6\endcsname{\color{black}}%
      \expandafter\def\csname LT7\endcsname{\color{black}}%
      \expandafter\def\csname LT8\endcsname{\color{black}}%
    \fi
  \fi
  \setlength{\unitlength}{0.0500bp}%
  \begin{picture}(7200.00,5040.00)%
    \gplgaddtomacro\gplbacktext{%
      \csname LTb\endcsname%
      \put(1254,1017){\makebox(0,0)[r]{\strut{}-1}}%
      \put(1254,1644){\makebox(0,0)[r]{\strut{}-0.8}}%
      \put(1254,2270){\makebox(0,0)[r]{\strut{}-0.6}}%
      \put(1254,2897){\makebox(0,0)[r]{\strut{}-0.4}}%
      \put(1254,3523){\makebox(0,0)[r]{\strut{}-0.2}}%
      \put(1254,4150){\makebox(0,0)[r]{\strut{} 0}}%
      \put(1254,4776){\makebox(0,0)[r]{\strut{} 0.2}}%
      \put(1512,484){\makebox(0,0){\strut{}-3}}%
      \put(2398,484){\makebox(0,0){\strut{}-2}}%
      \put(3285,484){\makebox(0,0){\strut{}-1}}%
      \put(4172,484){\makebox(0,0){\strut{} 0}}%
      \put(5059,484){\makebox(0,0){\strut{} 1}}%
      \put(5946,484){\makebox(0,0){\strut{} 2}}%
      \put(6832,484){\makebox(0,0){\strut{} 3}}%
      \put(484,2740){\rotatebox{90}{\makebox(0,0){\strut{}$\mathrm{Re}\left<GP^*\right>_{\mathcal{S}^2}$}}}%
      \put(4172,154){\makebox(0,0){\strut{}$\beta$}}%
    }%
    \gplgaddtomacro\gplfronttext{%
    }%
    \gplbacktext
    \put(0,0){\includegraphics{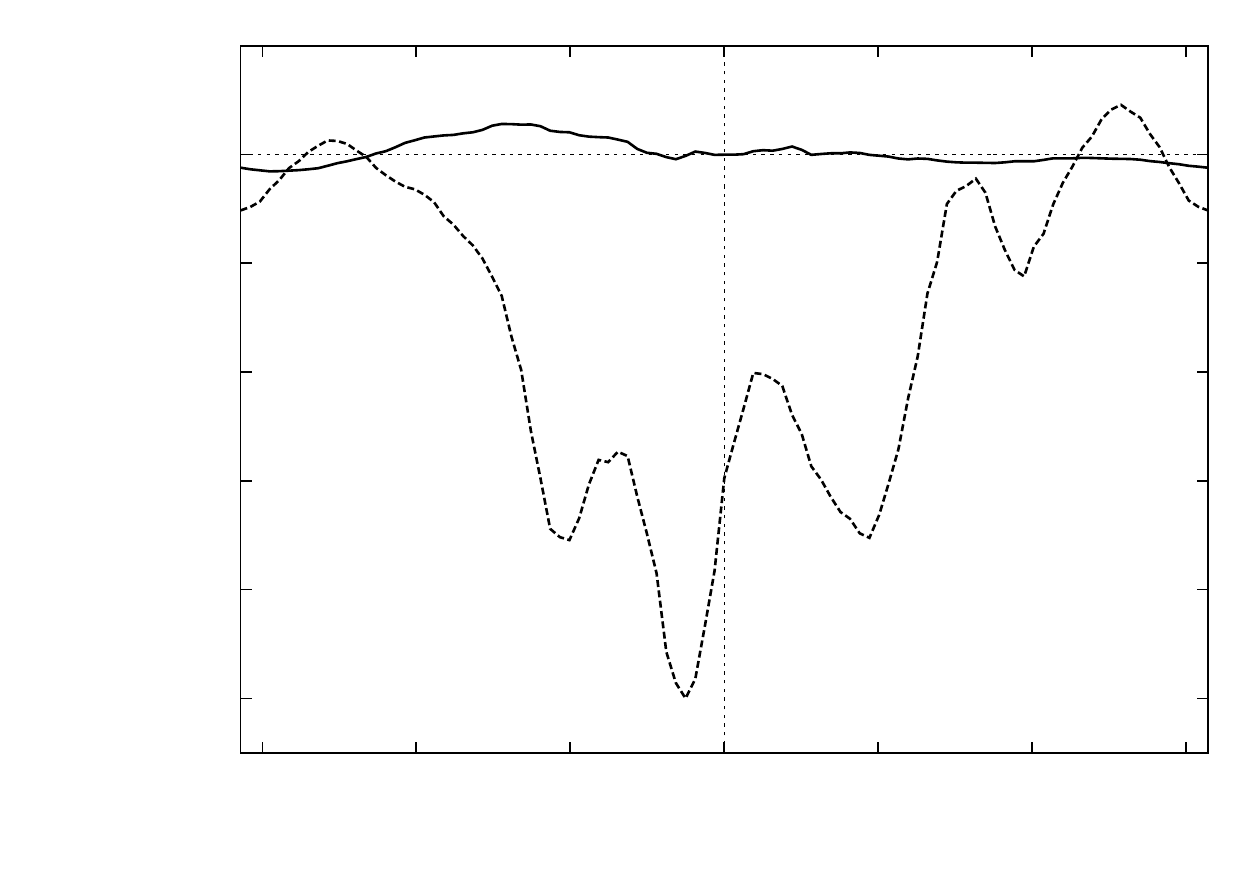}}%
    \gplfronttext
  \end{picture}%
\endgroup

%% file: rotation-WMAPa.tex
\begingroup
  \makeatletter
  \providecommand\color[2][]{%
    \GenericError{(gnuplot) \space\space\space\@spaces}{%
      Package color not loaded in conjunction with
      terminal option `colourtext'%
    }{See the gnuplot documentation for explanation.%
    }{Either use 'blacktext' in gnuplot or load the package
      color.sty in LaTeX.}%
    \renewcommand\color[2][]{}%
  }%
  \providecommand\includegraphics[2][]{%
    \GenericError{(gnuplot) \space\space\space\@spaces}{%
      Package graphicx or graphics not loaded%
    }{See the gnuplot documentation for explanation.%
    }{The gnuplot epslatex terminal needs graphicx.sty or graphics.sty.}%
    \renewcommand\includegraphics[2][]{}%
  }%
  \providecommand\rotatebox[2]{#2}%
  \@ifundefined{ifGPcolor}{%
    \newif\ifGPcolor
    \GPcolorfalse
  }{}%
  \@ifundefined{ifGPblacktext}{%
    \newif\ifGPblacktext
    \GPblacktexttrue
  }{}%
  \let\gplgaddtomacro\g@addto@macro
  \gdef\gplbacktext{}%
  \gdef\gplfronttext{}%
  \makeatother
  \ifGPblacktext
    \def\colorrgb#1{}%
    \def\colorgray#1{}%
  \else
    \ifGPcolor
      \def\colorrgb#1{\color[rgb]{#1}}%
      \def\colorgray#1{\color[gray]{#1}}%
      \expandafter\def\csname LTw\endcsname{\color{white}}%
      \expandafter\def\csname LTb\endcsname{\color{black}}%
      \expandafter\def\csname LTa\endcsname{\color{black}}%
      \expandafter\def\csname LT0\endcsname{\color[rgb]{1,0,0}}%
      \expandafter\def\csname LT1\endcsname{\color[rgb]{0,1,0}}%
      \expandafter\def\csname LT2\endcsname{\color[rgb]{0,0,1}}%
      \expandafter\def\csname LT3\endcsname{\color[rgb]{1,0,1}}%
      \expandafter\def\csname LT4\endcsname{\color[rgb]{0,1,1}}%
      \expandafter\def\csname LT5\endcsname{\color[rgb]{1,1,0}}%
      \expandafter\def\csname LT6\endcsname{\color[rgb]{0,0,0}}%
      \expandafter\def\csname LT7\endcsname{\color[rgb]{1,0.3,0}}%
      \expandafter\def\csname LT8\endcsname{\color[rgb]{0.5,0.5,0.5}}%
    \else
      \def\colorrgb#1{\color{black}}%
      \def\colorgray#1{\color[gray]{#1}}%
      \expandafter\def\csname LTw\endcsname{\color{white}}%
      \expandafter\def\csname LTb\endcsname{\color{black}}%
      \expandafter\def\csname LTa\endcsname{\color{black}}%
      \expandafter\def\csname LT0\endcsname{\color{black}}%
      \expandafter\def\csname LT1\endcsname{\color{black}}%
      \expandafter\def\csname LT2\endcsname{\color{black}}%
      \expandafter\def\csname LT3\endcsname{\color{black}}%
      \expandafter\def\csname LT4\endcsname{\color{black}}%
      \expandafter\def\csname LT5\endcsname{\color{black}}%
      \expandafter\def\csname LT6\endcsname{\color{black}}%
      \expandafter\def\csname LT7\endcsname{\color{black}}%
      \expandafter\def\csname LT8\endcsname{\color{black}}%
    \fi
  \fi
  \setlength{\unitlength}{0.0500bp}%
  \begin{picture}(7200.00,5040.00)%
    \gplgaddtomacro\gplbacktext{%
      \csname LTb\endcsname%
      \put(1254,704){\makebox(0,0)[r]{\strut{} 0}}%
      \put(1254,1444){\makebox(0,0)[r]{\strut{} 0.2}}%
      \put(1254,2185){\makebox(0,0)[r]{\strut{} 0.4}}%
      \put(1254,2925){\makebox(0,0)[r]{\strut{} 0.6}}%
      \put(1254,3665){\makebox(0,0)[r]{\strut{} 0.8}}%
      \put(1254,4406){\makebox(0,0)[r]{\strut{} 1}}%
      \put(1512,484){\makebox(0,0){\strut{}-3}}%
      \put(2398,484){\makebox(0,0){\strut{}-2}}%
      \put(3285,484){\makebox(0,0){\strut{}-1}}%
      \put(4172,484){\makebox(0,0){\strut{} 0}}%
      \put(5059,484){\makebox(0,0){\strut{} 1}}%
      \put(5946,484){\makebox(0,0){\strut{} 2}}%
      \put(6832,484){\makebox(0,0){\strut{} 3}}%
      \put(484,2740){\rotatebox{90}{\makebox(0,0){\strut{}$\mathrm{Re}\left<GP^*\right>_{\mathcal{S}^2}$}}}%
      \put(4172,154){\makebox(0,0){\strut{}$\beta$}}%
    }%
    \gplgaddtomacro\gplfronttext{%
    }%
    \gplbacktext
    \put(0,0){\includegraphics{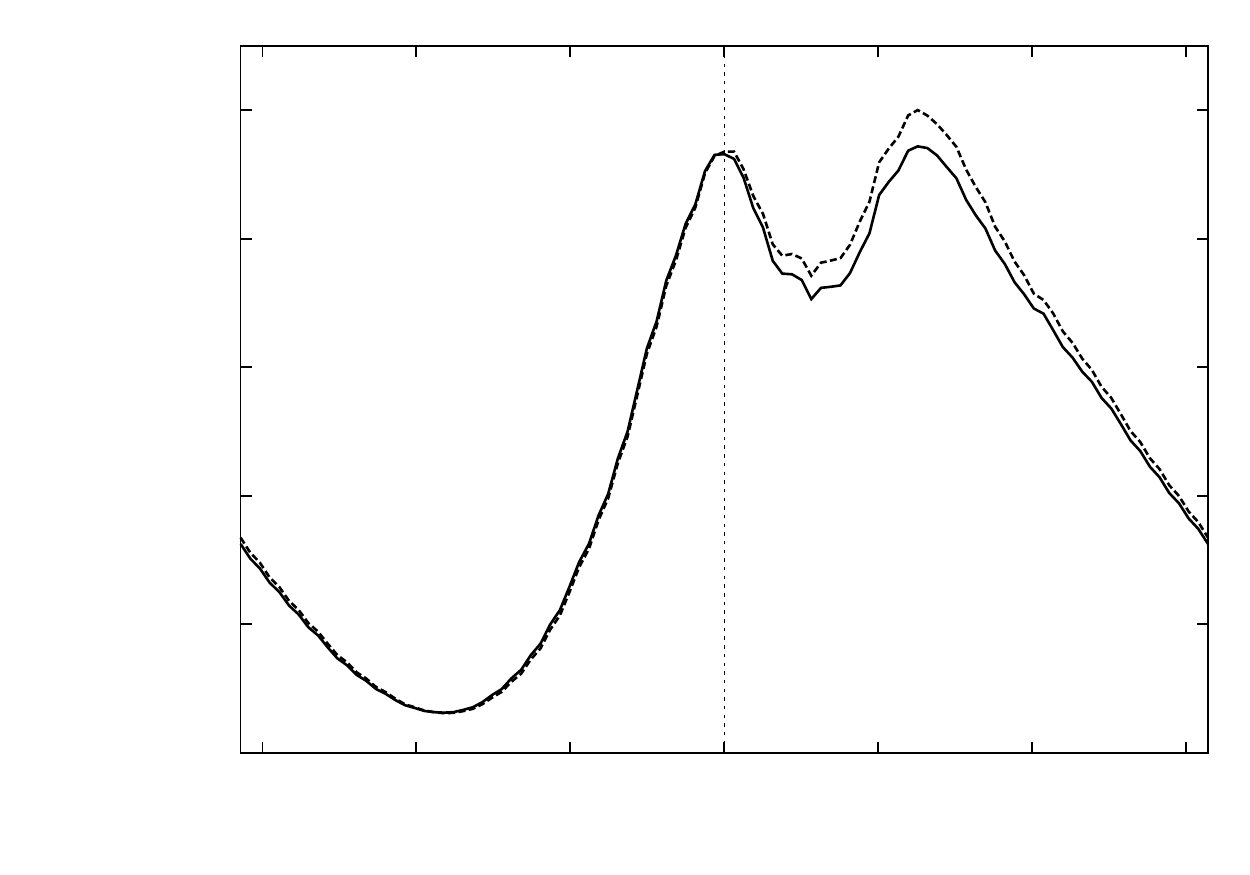}}%
    \gplfronttext
  \end{picture}%
\endgroup